\documentclass[]{jkas} 

\def\year{2023} 
\def\volume{56} 
\def\issue{1} 
\def\beginpage{1} 
\def\received{---} 
\def\accepted{---} 
\def\published{---} 

\date{Received \received; Accepted \accepted; Published \published}

\newcommand\msun{{\,M_\odot}}

\title{%
K-DRIFT Science Theme:\\
New Theoretical Framework Using the Galaxy Replacement Technique for LSB studies
}

\author[1,$\star$]{Kyungwon Chun}{0000-0001-9544-7021}
\author[1,2]{Jihye Shin}{0000-0001-5135-1693}
\author[3]{Rory Smith}{0000-0001-5303-6830}
\author[1,2]{Jongwan Ko}{0000-0002-9434-5936}
\author[1,4]{Jaewon Yoo}{0000-0002-6841-8329}
\author[1]{So-Myoung~Park}{0000-0003-1889-325X}
\author[1]{Woowon Byun}{0000-0002-7762-7712}
\author[1]{Sang-Hyun Chun}{0000-0002-6154-7558}
\author[1]{Sungryong Hong}{0000-0001-9991-8222}
\author[3,1]{Hyowon Kim}{0000-0003-4032-8572}
\author[1]{Jae-Woo Kim}{0000-0002-1710-4442}
\author[1,5]{Jaehyun Lee}{0000-0002-6810-1778}
\author[1,2]{Hong Soo Park}{0000-0002-3505-3036}
\author[1,6]{Jinsu Rhee}{0000-0002-0184-9589}
\author[1,2]{Kwang-Il Seon}{0000-0001-9561-8134}
\author[1,7]{Yongmin Yoon}{0000-0003-0134-8968}

\affil[1]{Korea Astronomy and Space Science Institute, Daejeon 34055, Republic of Korea}
\affil[2]{Department of Astronomy and Space Science, University of Science and Technology, Daejeon 34113, Republic of Korea}
\affil[3]{Departamento de F{\'i}sica, Universidad T{\'e}cnica Federico Santa Mar{\'i}a, Avenida Espa{\~n}a 1680, Valpara{\'i}so, Chile}
\affil[4]{Quantum Universe Center, Korea Institute for Advanced Study, Seoul 02455, Republic of Korea}
\affil[5]{School of Physics, Korea Institute for Advanced Study, Seoul 02455, Republic of Korea}
\affil[6]{Institut d’Astrophysique de Paris, Sorbonne Universit{\'e}, CNRS, UMR 7095, 98 bis bd Arago, 75014 Paris, France}
\affil[7]{Department of Astronomy and Atmospheric Sciences, Kyungpook National University, Daegu 41566, Republic of Korea}

\def\corrauthor{%
K. Chun, \email{kwchun@kasi.re.kr}
}

\def\runningauthor{%
Chun et al.
}

\def\runningtitle{%
K-DRIFT Science Theme: New Theoretical Framework using the GRT
}

\def\keywords{%
astronomy --- astrophysics --- journals: individual: JKAS
}

\def\abstracttext{%
Low-surface-brightness (LSB) structures provide critical insights into the hierarchical formation of galaxies and galaxy clusters. 
The KASI Deep Rolling Imaging Fast Telescope (K-DRIFT) is designed to detect such diffuse features through deep, wide-field optical imaging with a surface brightness reaching $\sim$$30~\rm{mag}~\rm{arcsec}^{-2}$. 
To interpret the observation data expected from K-DRIFT, we have developed the Galaxy Replacement Technique (GRT), an $N$-body simulation framework optimized for tracing the gravitational evolution of stellar components. 
The GRT works by inserting high-resolution galaxy models, including a dark matter (DM) halo and stellar disk, in place of multiple low-resolution DM halos in the base $N$-body cosmological simulation.
It allows us to achieve very high mass ($m_{star}=5.4\times10^4\msun\ h^{-1}$) and spatial resolution (10~$\rm{pc}~h^{-1}$) with shorter computation time compared to full hydrodynamic cosmological simulations.
Therefore, this technique is particularly well-suited for studying LSB structures, with a surface brightness reaching $\sim$$31~\rm{mag}~\rm{arcsec}^{-2}$.
In this paper, we present the motivation and methodology of the GRT, summarize key results from previous studies, and highlight its synergy with K-DRIFT observations. 
We further discuss planned science cases using the GRT, aiming to build a theoretical basis for interpreting LSB features in various environments.
}

\begin{document}
\jkashead 


\section{Introduction}
In the current concordance cosmology of Lambda cold dark matter ($\Lambda$CDM), structures such as galaxies, groups, and clusters are hierarchically built up by the merging of smaller structures \citep{press1974,fall1980,ryden1987,vbosch2002,agertz2011}.
Some of the accreted structures can survive as satellites, but some are also tidally disrupted.
In this process, low-surface-brightness (LSB) structures, such as diffuse light and tidal features, can form.
As these diffuse structures preserve signatures of past gravitational interactions, including mergers, stripping, and accretion events, they provide evidence of the hierarchical structure formation process \citep[][]{puchwein2010,montes2014,ko2018,pop2018,contini2019,mancillas2019}. 
Therefore, detecting and investigating LSB features are crucial to understanding the formation and evolution of structures on a cosmological scale.

Despite this importance, the study of LSB structures poses observational and theoretical challenges.
Observationally, the detection of LSB structures is fundamentally limited by systematic uncertainties.
Key challenges include contamination by stray light, inaccurate flat-fielding, and variations in the natural sky background.
To overcome these challenges, the Korea Astronomy and Space Science Institute (KASI) is developing the ``KASI Deep Rolling Imaging Fast Telescope" (K-DRIFT).
K-DRIFT is designed to achieve wide-field, clean, and flat imaging, enabling the detection of LSB features down to $30~\rm{mag}~\rm{arcsec}^{-2}$ (see \citealt{ko2025} for details).

From a theoretical side, cosmological simulations have provided valuable insights into LSB structures. To resolve diffuse features down to the K-DRIFT surface-brightness depth of $\mu \sim 30~\rm{mag}~\rm{arcsec}^{-2}$, simulations require sufficiently high stellar mass resolution($\sim 10^5~\msun$). High-resolution zoom-in simulations such as {\textsc NewHorizon2} \citep[][]{yi2024} and {\textsc NewCluster} \citep{han2026} can reach this surface brightness level, but their limited volumes contain only a few large-scale structures (e.g., groups and clusters), restricting statistical investigations. In contrast, large full-box simulations (e.g., Illustris TNG300; \citealt{nelson2019}, Horizon Run 5; \citealt{lee21}) can produce statistically meaningful structure samples, yet their limited mass and spatial resolution fail to resolve LSB features fainter than $29~\rm{mag}~\rm{arcsec}^{-2}$.


To overcome the limitations of the standard simulations, \cite{chun2022} introduced the Galaxy Replacement Technique (GRT).
The GRT is a novel multiresolution $N$-body simulation framework, which is designed and optimized for describing the gravitational evolution of stellar components in a hierarchical merging process.
As most LSB structures form by gravitational interaction between structures, the GRT does not include the hydrodynamical recipes that require many computational resources.
Instead, we insert high-resolution models of galaxies, including a DM halo and stellar disk, in place of low-resolution DM halos in cosmological $N$-body simulations.
It enables us to trace the spatial distribution and evolution of LSB structures in a short computational time. 
Thus, we can achieve enhanced mass and spatial resolution to model the tidal stripping process and the formation of LSB features.

The GRT has already been successfully applied to produce a statistically significant sample of 84 clusters \citep[][]{chun2023,chun2024,chun2026}, resolving LSB components like intracluster light (ICL) and tidal features such as tidal stream, tidal tail, and shell-like structures down to surface brightness limits of $\mu_{V}=31~\rm{mag}~\rm{arcsec}^{-2}$.
Thanks to the computational efficiency of the GRT, we have extended its application beyond clusters to a broader range of environments, including galaxy groups and Milky Way (MW)–mass halos.
This enables a study of the gravitational evolution of galaxies and LSB structures across different environments.
The suite of GRT simulations provides a robust theoretical counterpart to upcoming K-DRIFT observations, allowing for direct comparisons under detection conditions matched to those of K-DRIFT and offering valuable insights into the formation of observed structures by tracing their evolution on a cosmological scale.

As part of K-DRIFT white paper series, this paper is organized as follows. 
Section \ref{sec:framework} introduces K-DRIFT and the GRT framework. 
Sections \ref{sec:GRT2} and Section \ref{sec:case} highlight previous studies based on the GRT and present potential science cases. 
In Section \ref{sec:summary}, we summarize the synergies between K-DRIFT and the GRT.

\begin{figure*}
\centering
\includegraphics[width=0.9\textwidth]{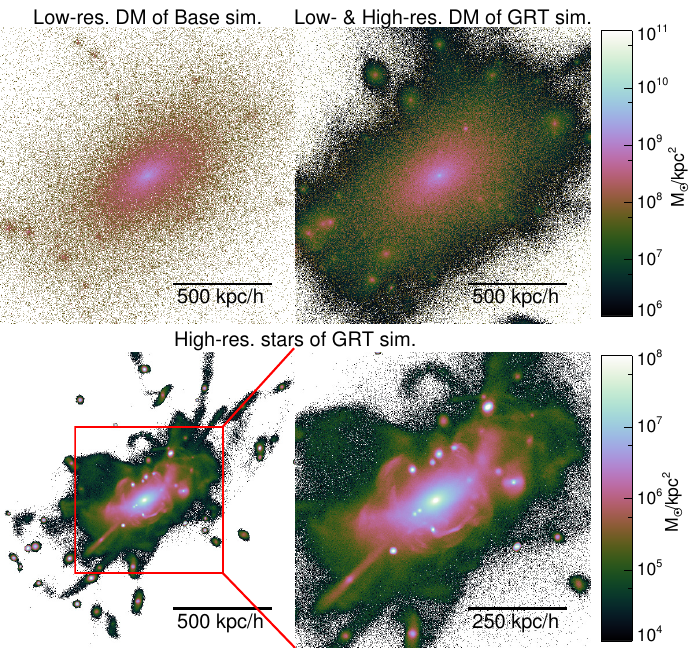}
\caption{DM and stellar surface mass density maps of the GRT cluster. The upper left and right panels show DM structures inside the virial radius ($R_{\rm{vir}}$) of the GRT cluster at $z=0$ in the base simulation and the GRT simulation, respectively. The lower left and right panels show the stellar structures in $R_{\rm{vir}}$ and 0.5$R_{\rm{vir}}$ of the GRT cluster. The color scale for DM and stellar structures is shown to the right of the panels. For reference, the stellar surface mass density range shown in the colorbar corresponds approximately to $\mu_V \sim 23.1-33.1~\rm{mag}~\rm{arcsec}^{-2}$ assuming a $V$-band mass-to-light ratio $(M_\star/L_V)$ of $5~M_\odot/L_{\odot,V}$. Figure credit from \cite{chun2022}.}
\label{fig:structure}
\end{figure*}

\section{GRT Simulations as a Framework for LSB Studies with K-DRIFT}\label{sec:framework}
\subsection{K-DRIFT: Deep imaging for faint structures}\label{sec:K-DRIFT}
K-DRIFT adopts a linear-astigmatism-free three-mirror-system \citep[LAF-TMS;][]{chang2025,park2020}.
\cite{byun2022} showed that this system is suitable for the observations of LSB structures through the test observations using the prototype of K-DRIFT, K-DRIFT Pathfinder, as it reaches $\mu_r\sim28.5~\rm{mag}~\rm{arcsec}^{-2}$.

To address issues identified during test observations with K-DRIFT Pathfinder and to achieve a wider field of view (FoV) optimized for LSB science, a new-generation telescope, K-DRIFT Generation 1 (G1), has been developed. 
K-DRIFT G1 consists of two identical telescopes with a sCMOS camera (Teledyne COSMOS-66) composed of an 8k $\times$ 8k single chip with a pixel scale of $\sim$$2''$, producing a wide FoV of $4.43^{\circ} \times 4.43^{\circ}$.
For K-DRIFT G1, Sloan $ugr$ and Luminance ($L$) filters are available.
The goal of the survey is to cover the area of declination from $-20^{\circ}$ to $-40^{\circ}$, achieving an imaging depth of $\sim$$30~\rm{mag}~\rm{arcsec}^{-2}$ in $10''\times10''$ boxes.
More details are available in \cite{ko2025}.

The K-DRIFT pathfinder observations have already revealed detectable tidal feature (see \citealt{byun2022,byun2026}). Mock images generated from GRT simulations can be used to interpret the visibility and morphology of these observed LSB features. However, the current pathfinder sample size is still limited for robust quantitative statistical comparisons, which will become possible with the substantially larger dataset expected from the upcoming K-DRIFT G1 survey.

\begin{figure}
\centering
\includegraphics[width=0.45\textwidth]{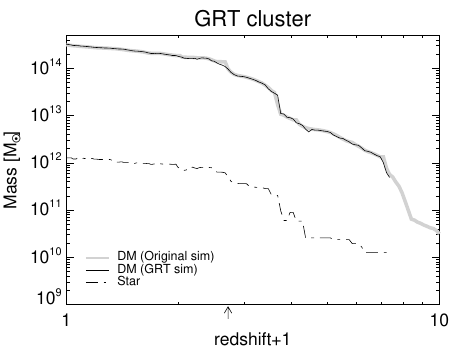}
\caption{Mass growth history of the GRT cluster. The thick gray and thin black solid lines represent the growth history of the cluster in the base simulation and the GRT simulation, respectively. The dash-dotted line indicates the mass growth of the stars in the GRT cluster. The black arrow below the x-axis means the epoch of the last major merger of the cluster. Figure credit from \cite{chun2022}.}
\label{fig:mass_growth}
\end{figure}

\subsection{GRT: A simulation approach}\label{sec:GRT}
This subsection is primarily based on \cite{chun2022}.
The GRT is a simulation technique that focuses specifically on galaxy interactions and on the formation and evolution of LSB structures, such as tidal features and diffuse light, while following stellar particles without including computationally expensive baryonic physics.
As a result, this approach allows us to trace the spatial distribution and evolution of LSB structures, in contrast to semi-analytic models (SAMs), which do not treat stars as individual particles.

To perform simulation with the GRT, we first perform the low-resolution DM–only cosmological simulations, named the $N$-cluster run, which consist of 64 uniform boxes each of volume (120~Mpc$h^{-1}$)$^3$.
The simulations are performed from $z=200$ to 0 using the Gadget-3 code \citep{springel2005} with $512^3$ particles per box, corresponding to a DM particle mass of $10^9~\msun~h^{-1}$ and a fixed comoving gravitational softening length of 2.3~kpc~$h^{-1}$.
For this simulation set, we use the post-Plank cosmological model of $\Omega_{m}=0.3$, $\Omega_{\Lambda}=0.7$, $\Omega_{b}=0.047$, and $h=0.684$.
This simulation set is used in many works \citep[e.g.,][Park et al., in prep.]{chun2022,jhee2022,kim2022,smith2022a,smith2022b,yoo2022,awad2023,chun2023,chun2024,dong2024,kim2024,chun2026}.

The initial condition for particle positions and velocities is generated by a MUSIC package\footnote{https://bitbucket.org/ohahn/music/} \citep{hahn2011}.
The halo and subhalo structures are identified with the modified 6D phase-space halo finder ROCKSTAR\footnote{https://bitbucket.org/pbehroozi/rockstar-galaxies} \citep{behroozi2013b}.
The lower mass limit of a halo is $\sim$$2\times10^{10}~\msun~h^{-1}$ ($N_{\rm{DM}}=20$).
The merger tree of these halos is built using Consistent Trees\footnote{https://bitbucket.org/pbehroozi/consistent-trees} \citep{behroozi2013c}.
From the merger tree, we identify the mass growth history of all halos that will later constitute the interesting target and its subhalos, regardless of whether the halos survive until $z=0$ or not.

Because the box size of the $N$-cluster run is large enough, on average, more than 50 cluster-mass halos ($M_{\rm{200c}} > 10^{14}\msun~h^{-1}$ at $z=0$), 900 group-mass halos ($10^{13}\msun~h^{-1} < M_{\rm{200c}} < 10^{14}\msun~h^{-1}$ at $z=0$), and 8,000 massive galaxies ($10^{12}\msun~h^{-1} < M_{\rm{200c}} < 10^{13}\msun~h^{-1}$ at $z=0$) naturally form within each box, where $M_{\rm{200c}}$ is enclosed mass within a radius where the density equals 200 times the critical density of the universe at a specific redshift.
In this section, we present results for a single example cluster, referred to as the `GRT cluster,' to demonstrate the performance of the GRT.
Previous studies and planned science cases based on larger samples will be discussed in Sections \ref{sec:GRT2} and Section \ref{sec:case}.

From the merger tree, we identify all the halos that will later contribute to the mass growth of the GRT cluster and its satellites.
To trace the evolution of the GRT cluster with higher resolution particles, we replace each low-resolution DM halo with a high-resolution model that consists of a high-resolution DM halo and a stellar disk composed of high-resolution star particles.
In this process, we only replace halos with $M_{\rm{peak}} > 10^{11}\msun~h^{-1}$ that consist of more than 100 low-resolution DM particles, where $M_{\rm{peak}}$ is the maximum mass of a DM halo before falling into a more massive halo.
\cite{chun2022} showed that halos with $M_{\rm{peak}} < 10^{11}\msun~h^{-1}$ contribute only $\sim2\%$ of the total stellar mass in clusters at $z=0$; therefore, we do not replace them to reduce the computational cost.
As a result, tidal features originating from the disruption of these low-mass galaxies are not modeled, which may lead to an underestimation of the faintest tidal features.
However, given the surface-brightness depth targeted by K-DRIFT ($\mu \sim 30~\mathrm{mag\,arcsec^{-2}}$), such features are expected to be detectable only in nearby systems.
Moreover, although interactions between low-resolution halos and high-resolution galaxies could induce numerical heating, \cite{chun2022} demonstrated that such effects do not lead to artificial disruption of resolve galaxies.
Therefore, this replacement threshold does not significantly limit meaningful statistical comparisons between GRT and K-DRIFT.

A halo is replaced with the high-resolution model when one of the following criteria is first satisfied: (1) the halo reaches $M_{\rm{peak}}$, (2) the halo first accretes a replaced satellite.
In the high-resolution model, we assume the Navarro-Frank-White (NFW) density profile \citep{navarro1997} for the DM halo and a bulgeless exponential disk galaxy for the stellar disk, respectively.
This is reasonable, as most low-mass disk galaxies with $M_{\rm{star}} < 10^{10}\msun$ have this morphology \citep{dutton2009}.
The mass of the stellar disk is determined using the redshift-dependent stellar-to-halo mass relation of \cite{behroozi2013a}. \footnote{In the replacement process, unbound particles within the virial radius are removed, whereas the halo mass is defined using only bound particles. Although this could make a small difference in the stellar mass assigned to the replaced halos, we find that the fraction of unbound particles within the virial radius at the replacement epoch is small (typically $\sim2\%$). Therefore, this effect does not significantly affect the robustness of our method.}
The scale length of the stellar disk is determined using the mass-size relation of observed galaxies \citep{dutton2011}, and the scale height is fixed at 15\% of the scale length.
Because our simulation is a collisionless N-body run, we assign a formation epoch and metallicity to each stellar particle following the prescriptions described in \cite{chun2024}. Briefly, the formation epoch is derived from the stellar mass growth history inferred from the halo mass assembly using the stellar-to-halo mass relation of \cite{behroozi2013a}, while the metallicity of newly formed stars is estimated from the mass–metallicity relation of \cite{ma2016}. 
We assume a radial gradient such that inner-disk particles have higher metallicity and younger ages.
The stellar luminosity evolution is computed using the stellar population synthesis model of \cite{bruzual2003}.
Even though we consider only disk galaxies in our high-resolution model, we note that bulges, spheroids, and other dispersion-supported structures naturally evolve in our simulation as galaxies merge and interact tidally.

In order to complete the GRT simulation, we perform our multiresolution resimulation until $z=0$.
The GRT simulation has a gravitational softening length for the high-resolution DM and star particles of $\sim$100 and $\sim$$10~$pc$~h^{-1}$, respectively.
With such a high spatial resolution, we can well resolve the self-gravity of the disks radially and vertically out of the disk plane and accurately follow the tidal stripping process.
The gravitational softening length for the low-resolution DM is $2.3~\rm{kpc}~h^{-1}$, which is the same as in the base simulation ($N$-cluster run).
The high-resolution particle mass for DM and star is $5.4\times10^6\msun~h^{-1}$ and $5.4\times10^4\msun~h^{-1}$, respectively.

The mass of the high-resolution star particles is similar to the baryon mass of the minimum mass of the star particles in Illustris TNG50 or \textsc{NewHorizon} simulations \citep{nelson2019,dubois2021}.
However, those two different simulations used 130 million and 40 million CPU hours to simulate until $z=0$ and $z=0.7$, respectively. 
Note that there is only one cluster in Illustris TNG50 and one group in NewHorizon, so it is impossible to consider the influence of cluster-to-cluster variations on the LSB features in the clusters using these simulations.
On the other hand, our GRT simulation used only 50,000 CPU hours to trace the evolution of one cluster up to $z=0$.
Thus, the GRT opens up the possibility of conducting large statistical studies of the LSB structures with a live distribution of star particles in groups and clusters for the first time.

Figure \ref{fig:structure} shows DM and stellar structures of the GRT cluster, colored by surface mass density.
Clearly, the surface density where we can make reliable predictions about the DM structures is much lower ($<$$10^8\msun~$kpc$^{-2}$) in the GRT simulation than in the base simulation (compare the top left and right panels).
The stellar structures of the GRT cluster, as shown in the bottom left panel, reveal the diffuse light and tidally disrupted structures surrounding the brightest cluster galaxy (BCG) at the center.
The bottom right panel shows the complex structures in the central region with a zoom-in view for more detail.

Additionally, Figure \ref{fig:mass_growth} shows the mass growth history of the GRT cluster, and we find that the mass evolution of the GRT cluster follows that of the base simulation well.
As the stars are supplied by the high-resolution galaxies that fall into the cluster, the stellar mass growth (dashed-dotted line) shows similar growth episodes to the DM mass growth.

In summary, the GRT focuses on accurately modeling large samples of galaxies, groups, and clusters with resolved LSB features in a full cosmological context.
As the GRT is a multiresolution cosmological $N$-body resimulation, this allows us to place the high-resolution where it is most needed at low computational cost.
This is particularly important because typical large-volume hydrodynamical cosmological simulations (e.g., Illustris TNG100, EAGLE, etc.) have a spatial resolution of $\sim$1~kpc, which cannot resolve the small disks ($\sim$1~kpc) of low-mass galaxies.
Indeed, \cite{genel2018} showed that their model galaxies ($M_{\rm{star}} > 10^9\msun$) tend to have larger disks compared to observations, which makes them unrealistically susceptible to tidal stripping \citep{smith2016}.
This will eventually affect the growth and evolution of more massive galaxies that form hierarchically from them.
By contrast, with our high spatial resolution and precise control over the size of the stellar disks that we insert, our galaxies are more realistically robust to external tides \cite[see Section 3.2 of][for more details]{chun2022}.
Moreover, the low mass of star particles ($5.4\times10^4\msun~h^{-1}$) allows for reliable modeling of faint LSB features ($\sim$$31~$mag~arcsec$^{-2}$).
The use of abundance matching also ensures that our galaxy masses match the observed galaxy masses over a broad range of redshift.

In conclusion, the GRT will provide a powerful theoretical tool for upcoming LSB surveys.
Its design is particularly synergistic with the goals of the K-DRIFT project.
By providing cosmologically based predictions of LSB features, the simulations with the GRT give a critical theoretical basis for interpreting K-DRIFT observations.
A series of recent studies using the GRT \citep[][]{chun2022,yoo2022,chun2023,chun2024,chun2026} have shown the potential of the GRT to uncover the physical origin of LSB structures across various environments.
We will highlight the main results from these works in Section \ref{sec:GRT2}.

\section{LSB studies using GRT simulations}\label{sec:GRT2} 
This section is primarily based on our works \citep[][]{chun2022,chun2023,chun2024,chun2026}.
The GRT, introduced in Section \ref{sec:GRT}, provides a powerful numerical framework for studying the formation and evolution of LSB structures, including tidal tails, tidal streams, shell-like structures, and diffuse light.
In this section, we summarize the key results obtained from the analysis of 84 GRT clusters ($10^{13.6} < M_{\rm{200c}} [\msun] < 10^{14.8}$) in our studies \citep[][]{chun2022,chun2023,chun2024,chun2026}.

\begin{figure}
\centering
\includegraphics[width=0.45\textwidth]{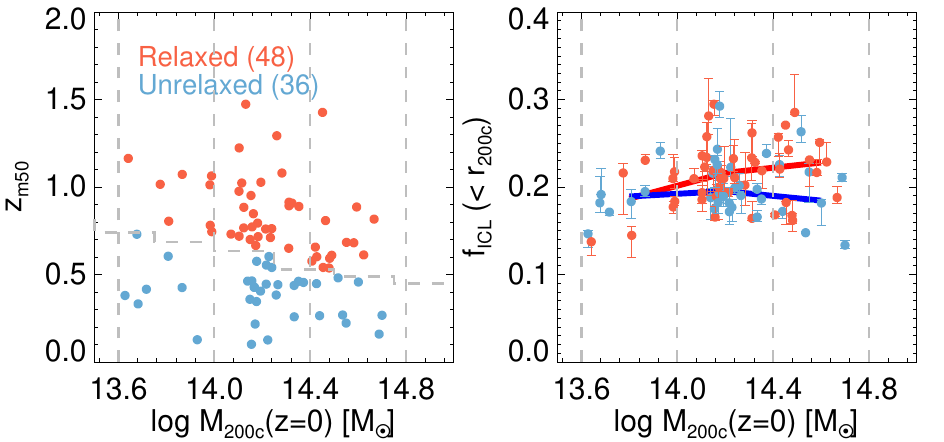}
\caption{Left: the relation between $M_{\rm{200c}}$ and $z_{\rm{m50}}$ of GRT clusters at $z=0$. The filled red and blue circles indicate the relaxed and unrelaxed clusters. The horizontal dashed gray line is the median $z_{\rm{m50}}$ of all clusters in the $N$-cluster run within a logarithmic mass bin of 0.25 dex. Right: the relation between $M_{\rm{200c}}$ and ICL fraction ($f_{\rm{ICL}}$) within $r_{\rm{200c}}$ of GRT clusters at $z=0$. The filled red and blue circles indicate the median $f_{\rm{ICL}} (< r_{\rm{200c}})$ among the ICL fractions calculated on the three different planes for relaxed and unrelaxed clusters. The error bar caps are the upper and lower values of the ICL fractions for each cluster. Figure adapted from \cite{chun2024}.}
\label{fig:ficl}
\end{figure}

Before describing the key results, we briefly present the basic properties of 84 GRT clusters. The left panel of Figure \ref{fig:ficl} shows the mass distribution of 84 GRT clusters and their $z_{\rm{m50}}$ parameter to quantify the mass growth history of each cluster and to represent the dynamical state of each cluster.
Indeed, previous simulation studies have shown that clusters with higher $z_{\rm{m50}}$ are more dynamically relaxed \citep[e.g.,][]{power2012,mostoghiu2019,haggar2020,chun2023}.
To classify the relaxed and unrelaxed clusters among GRT clusters, we compute the median value of $z_{\rm{m50}}$ within a logarithmic mass bin of 0.25~dex using all clusters within a mass range of $M_{\rm{200c}}=10^{13-15}\msun$ in the $N$-cluster run.
We classify clusters with $z_{\rm{m50}}$ higher than the median value of $z_{\rm{m50}}$ as relaxed clusters and others as unrelaxed clusters.
This panel shows that the $z_{\rm{m50}}$ parameter decreases as the mass of clusters increases due to the hierarchical mass growth in $\Lambda$CDM cosmology.
We again emphasize that such a large cluster sample, resolved at high spatial($\sim$$10~$pc$~h^{-1}$) and mass($5.4\times10^4\msun~h^{-1}$) resolutions, is achievable thanks to the computational efficiency of the GRT.
This is a significant advantage for statistical studies of LSB structures, especially when compared to high-resolution hydrodynamical simulations such as TNG50, which achieve comparable resolution but contain only a single galaxy cluster.

\subsection{Fraction of diffuse light} \label{sec:frac}
In the $\Lambda$CDM model, the galaxy groups and clusters grow their mass by merging with many smaller structures.
These smaller structures gravitationally interact with other structures in the cluster (or group) and lose their components or even become completely disrupted. 
In this process, many stellar components, among them, are spread throughout the cluster (or group) and evolve as diffuse light.
This diffuse light is referred to as the intragroup light (IGL) within the group and the ICL within the cluster, respectively \citep[for a review, see ][]{contini2021,montes2022}.
Throughout this section, we use the term ``ICL’’ to refer to diffuse light in both galaxy groups and clusters for simplicity.
The ICL components are defined as the stellar components in regions fainter than $\mu_{V}~>~26.5~\rm{mag}~\rm{arcsec}^{-2}$ in a 2D surface brightness map.

The right panel of Figure \ref{fig:ficl} shows the ICL fraction ($f_{\rm{ICL}}$) of GRT clusters at $z=0$, defined as the fraction of the ICL luminosity to the total luminosity within $r_{\rm{200c}}$ of clusters. 
Although the ICL is ubiquitous in evolved massive structures, there is considerable variation in the fraction of ICL to the total luminosity in GRT clusters (10--30\%).
This diversity in ICL fractions has indeed been shown in previous observations \cite[e.g.,][]{burke2015,morishita2017,jimenez-teja2018,montes2018,spavone2018,furnell2021,poliakov2021,ragusa2021,yoo2021,joo2023}.

Despite the diversity in ICL fractions among GRT clusters, we find that the median ICL fraction for relaxed clusters is higher than that for unrelaxed clusters.
This difference suggests that the dynamical state of clusters may play a role in the ICL fraction of clusters, as both observations and theoretical studies have suggested \cite[e.g.,][]{darocha2008,montes2018,ragusa2021,ragusa2023,contini2023,contini2024,contreras2024}.
However, other studies have shown an opposite trend in which unrelaxed clusters show higher ICL fractions \citep[e.g.,][]{jimenez-teja2018,jimenez-teja2024}, motivating the need for a large and uniformly analyzed cluster sample.
The K-DRIFT wide-deep survey plan will cover approximately 1,100 galaxy clusters and may potentially extend to 4,300 clusters \citep{yoo2025}.
For these clusters, we will measure the ICL fraction through a consistent definition and uniform observation facilitated by a single instrument.
Thanks to the efficient and accurate modeling of the ICL with the GRT, we can also perform the GRT simulations targeting galaxy clusters with broad masses and dynamical states.
By applying the same observational limits to the GRT simulations, we can make ICL maps that are directly comparable to K-DRIFT observations. 
This will allow us to interpret the observed diversity in ICL fractions and their correlations with dynamical state.

\subsection{Formation channels of diffuse light}

\begin{figure}
\centering
\includegraphics[width=0.45\textwidth]{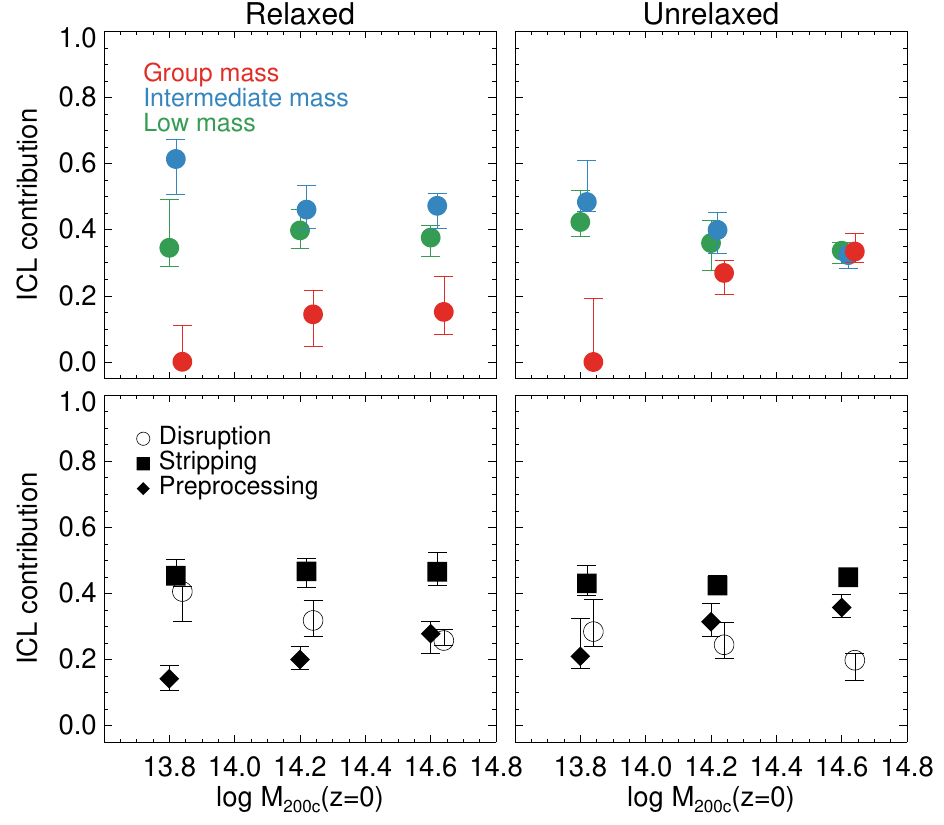}
\caption{Top: The contribution to the total ICL luminosity within the relaxed and unrelaxed GRT clusters depending on the mass of the satellites. The filled red, blue, and green circles indicate the subgroups called the ``group mass,’’ ``intermediate mass,’’ and ``low mass’’ satellites, and each error bar cap means the first and third quartiles for each subgroup. Bottom: The relation between the mass of clusters and the contribution of satellites to the ICL within the clusters, categorized depending on the three different formation mechanisms. The open circles, filled squares, and filled diamonds indicate the contribution of disrupted ICL, stripped ICL, and preprocessed ICL to the total ICL stars. Figure adapted from \cite{chun2024}.}
\label{fig:ficl_type}
\end{figure}

As the satellite galaxies gravitationally interact with other structures in the cluster, they lose some parts of their stellar components or can even be completely disrupted.
This transformation of satellites naturally generates the ICL within the cluster \citep[e.g.,][]{zibetti2005,conroy2007,murante2007,burke2015}.

Using the GRT simulations with 84 clusters, \cite{chun2024} showed that the ``intermediate mass’’ satellites ($10^{10} < M_{\rm{gal}} [\msun] < 10^{11}$) are the dominant contributors to the ICL across a wide range of cluster masses and dynamical states (the upper panel of Figure \ref{fig:ficl_type}).
On the other hand, ``group mass’’ satellites ($M_{\rm{gal}} [\msun] > 10^{11}$) tend to contribute more significantly to the ICL in more massive and unrelaxed clusters.
The importance of group infall in forming the ICL in the dynamical young clusters has indeed been observed in the Virgo cluster \citep{mihos2005,mihos2017}.
This is attributed to the presence of preprocessed ICL stars.
Indeed, \cite{chun2024} demonstrated that 50-60\% of ICL stars originating from the ``group mass’’ satellites enter the ICL regime before the satellites entered the cluster, i.e., preprocessing.
As a result, the relative contribution of preprocessed ICL stars increases in more massive and unrelaxed clusters (the bottom panel of Figure \ref{fig:ficl_type}).
Furthermore, \cite{chun2023} showed that, in the unrelaxed clusters, the contribution of preprocessed ICL stars to the outer ICL ($>$$0.5~R_{\rm{vir}}$) increases as these stars can be easily stripped from their host progenitors (see Figure 7 of \citeauthor{chun2023}~\citeyear{chun2023}).
Nonetheless, \cite{chun2024} concluded that tidal stripping within the cluster is the main formation mechanism of the ICL, regardless of the mass or dynamical state of the cluster (the bottom panels of Figure \ref{fig:ficl_type}).
This is consistent with previous studies \cite[e.g.,][]{rudick2009,puchwein2010,montes2014,contini2018,ragusa2021}.
However, the dominant formation channel of the ICL remains under active discussion, as some recent studies have suggested the alternative dominant channels \citep[e.g.,][]{joo2025,jeon2026}.

From an observational perspective, the radial color gradient of the ICL provides a clue to its formation mechanisms \citep{ko2018,joo2023}.
For example, a negative color gradient suggests ICL formation via tidal stripping of satellite galaxies.
In contrast, a flat or negligible color gradient would indicate the formation of ICL by major merger.
The K-DRIFT and GRT will be well-suited to test these predictions.
With observation of thousands of clusters ($z<0.1$), K-DRIFT will be able to measure the radial color gradient of the ICL.
In parallel, because the luminosity evolution of each stellar particle in the GRT simulations is computed using the stellar population synthesis model of \cite{bruzual2003}, the mock clusters enable a direct investigation of the expected ICL color gradient generated by different formation mechanisms.
As the mock images can be generated by applying observational definitions and measurement techniques consistent with K-DRIFT, we can perform a comparison between theory and observation.
Therefore, these will provide critical constraints on the dominant ICL formation mechanisms in the observed clusters.

\begin{figure*}
\centering
\includegraphics[width=0.95\textwidth]{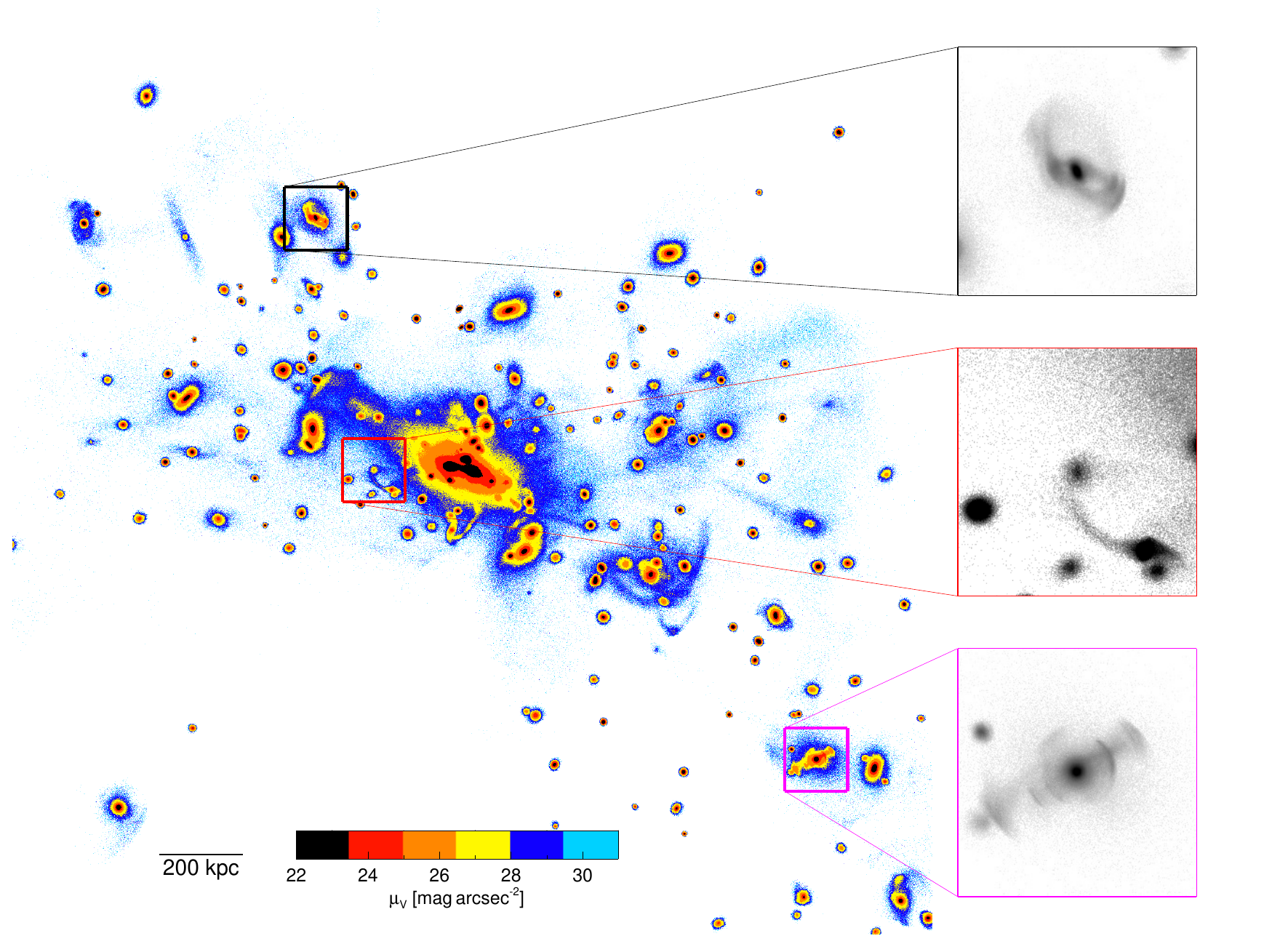}
\caption{$V$-band surface brightness ($\mu_V$) map of the most massive GRT cluster at $z=0$. Color bars indicate the $V$-band surface brightness. The zoomed-in images on the right side show the stellar distribution diagram of the representative tidal tail, stream, and shell-like structures. Figure adapted from \cite{chun2026}.}
\label{fig:map}
\end{figure*}

\subsection{Tidal features in clusters}

\begin{figure*}[!t]
\centering
\includegraphics[width=0.9\textwidth]{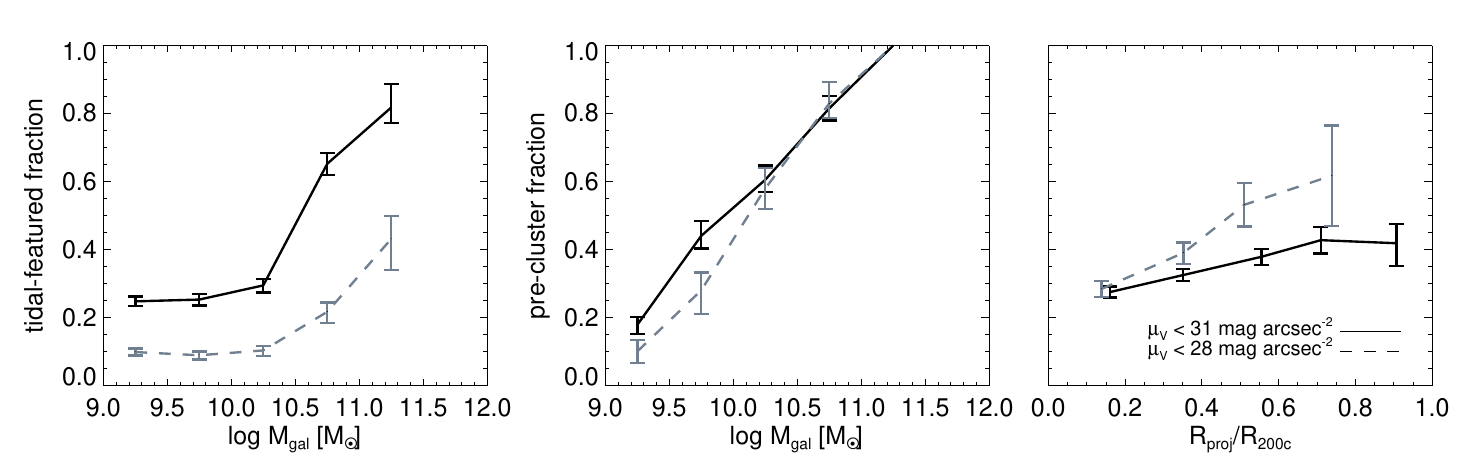}
\caption{The radial profile of tidal-featured fraction and the pre-cluster fraction in the cluster. In each panel, the solid and dashed lines represent all tidal-featured galaxies brighter than 31~mag~arcsec$^{-2}$ and 28~mag~arcsec$^{-2}$. The error bars represent the 1$\sigma$ uncertainties derived from bootstrap sampling. The bins containing fewer than five galaxies are omitted. Figure adapted from \cite{chun2026}.}
\label{fig:precluster}
\end{figure*}

Many numerical studies have shown that details of the tidal features, such as morphological characteristics, prominence, and the number, provide hints about the mass ratio of the merger, the orbital information, and the time since the last merging events \citep[e.g.,][]{cooper2010,sanderson2010,cooper2011,sanderson2013,ji2014,amorisco2015,pop2018,khalid2024}.

\cite{chun2026} studied how tidal features\footnote{In \cite{chun2026}, tidal features consist of unbounded stellar components classified by the internal dynamics of each galaxy with $M_{\rm{gal}}~>~10^9~\msun$ rather than visual inspection. See Section 2.2 of their manuscript for more details.} of galaxies form and evolve in and around cluster environments using GRT simulations.
Figure \ref{fig:map} shows a $V$-band surface brightness map of stars located in a 2~Mpc cube centered on the most massive GRT cluster ($5\times10^{14}\msun$).
Three squares indicate representatives of the tidal-featured galaxies.
The intracluster region, shown in yellow, blue, cyan in the figure, is filled with stars in the LSB regime of $\mu_{V}~>~26.5~\rm{mag}~\rm{arcsec}^{-2}$, while finer structures originate from tidal features, such as tail, stream, and shell.
The right zoomed-in images show the stellar distribution near tidal-featured galaxies.
Note that, due to the definition, the tidal features in \cite{chun2026} only include the tidal stream, tail, and shell-like structures, but do not include bounded tidal features (tidal-induced bar, strong spiral, asymmetric halo, double nucleus, etc.).

Among the 3262 cluster galaxies with $M_{\rm{gal}}~>~10^9~\msun$ in the 84 GRT clusters, 985 are classified as tidal-featured galaxies.
Although the strong tidal field in the cluster environment accelerates the mass loss of the satellites, only 20\% of tidal-featured galaxies form their tidal features purely due to tidal perturbation in the cluster environment, without any interaction with other galaxies with $M_{\rm{gal}}~>~10^7~\msun$ before falling into the cluster.
In contrast, the majority (80\%) of the tidal-featured galaxies are affected by the preprocessing, as they experienced merging events with other galaxies before falling into the cluster.
Specifically, 45\% of the tidal-featured galaxies form their tidal features after passing the pericenter of the cluster, affected by both preprocessing and the tidal field of the cluster, but 35\%, referred to as pre-cluster galaxies, form their tidal features before reaching the pericenter, primarily due to preprocessing.
Although a different detection method (i.e., visual inspection) may change the identified population of satellites with tidal features, the authors find that the tidal-featured galaxies that experienced preprocessing are still the dominant populations (68\%).

Additionally, \cite{chun2026} showed a higher tidal-featured fraction for the massive galaxies (the left panel of Figure \ref{fig:precluster}).
This trend is also evident in a brighter surface brightness limit (the dashed line; $\mu_{V}~<~$28~mag~arcsec$^{-2}$) and can be explained by the increase in the pre-cluster tidal-featured galaxies, as shown in the middle panel.
When we confine the tidal-featured galaxies to those identified with a brighter tidal feature than 28~mag~arcsec$^{-2}$, the tidal-featured fraction significantly decreases compared to that of 31~mag~arcsec$^{-2}$.
On the contrary, the pre-cluster fraction with $\mu_{V}~<~$28~mag~arcsec$^{-2}$ significantly increases (the right panel of Figure \ref{fig:precluster}).
This increase becomes more prominent in the outer regions, and paradoxically, suggests that deeper observations could reveal many more faint tidal-featured galaxies whose tidal features formed within the cluster environment.

In summary, the results highlight not only that preprocessing boosts the formation of tidal features and is the dominant path for forming the tidal features of satellites in clusters, but also that current observations of tidal-featured galaxies may be more biased toward pre-cluster galaxies.
As K-DRIFT will be able to detect faint and extended tidal features in galaxy clusters with deep photometric depth and wide-area coverage extending to the outer regions of clusters, this will enable a robust comparison between observations and theories, combined with physically motivated classifications from GRT simulations.
This synergy will not only help identify the diversity of tidal-featured galaxies but also provide more information about their formation and evolution.

\subsection{The Effect of the Observational Limits on the LSB structures}\label{limit}
The LSB structures, such as tidal features and diffuse light, contain valuable information about the hierarchical growth and interaction histories of galaxies and clusters. 
However, the detectability of these structures is strongly influenced by observational limitations, including surface brightness limit and FoV.
These factors can lead to an underestimation of the luminosity and spatial extent of LSB features, or even miss their presence.
In this section, using the GRT simulations, we examine how observational limitations affect the study of LSB structures.

\begin{figure}
\centering
\includegraphics[width=0.45\textwidth]{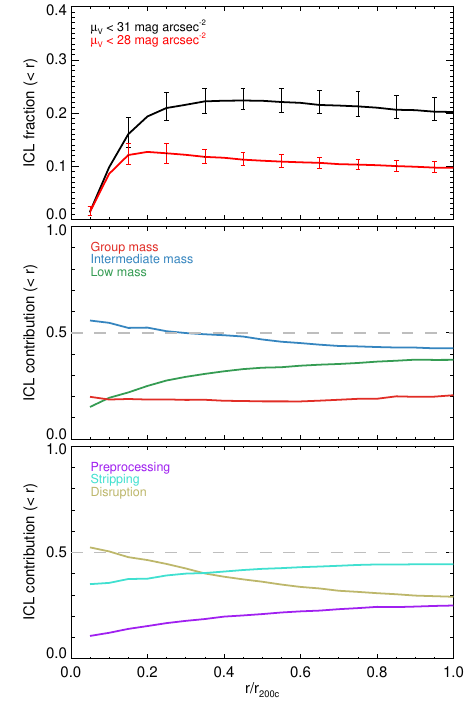}
\caption{Cumulative luminosity fraction of ICL stars and the contributions from different formation channels as a function of distance from the BCG, for the GRT clusters with $10^{14.0} < M_{\rm{200c}} [\msun] < 10^{14.4}$. The distance is normalized by $r_{\rm{200c}}$. In the top panel, the solid black and red lines indicate the radial profiles of ICL fractions down to $\mu_{V}=31~\rm{mag}~\rm{arcsec^{-2}}$ and $\mu_{V}=28~\rm{mag}~\rm{arcsec^{-2}}$, respectively, at $z=0$. Error bars represent the first and third quartiles of ICL fractions and contributions at each radius. The bottom two panels show the relative contributions from subgroups categorized by progenitor mass and formation mechanism, using the surface brightness limit of $\mu_{V}=31~\rm{mag}~\rm{arcsec^{-2}}$. Figure adapted from \cite{chun2024}.}
\label{fig:profile}
\end{figure}

Previous studies have highlighted the recoverability of faint LSB structures through the application of a lower faint-end surface brightness limit, which enables deeper observations \citep[e.g.,][]{burke2015,mihos2017,martin2022}.
\cite{burke2015}, in particular, emphasized the necessity of quantifying the impact of the faint-end surface brightness limit on the ICL fraction due to its strong dependence on this limit.
The top panel of Figure \ref{fig:profile} illustrates the cumulative luminosity fraction of ICL stars within the clusters of $10^{14.0} < M_{\rm{200c}} [\msun] < 10^{14.4}$ as a function of distance from the BCG, normalized by $r_{\rm{200c}}$.
In the panel, we can see that a higher faint-end surface brightness limit ($\mu_{V}=28~\rm{mag}~\rm{arcsec^{-2}}$) leads to an underestimation of the ICL fraction by about 40\% compared to the case with $\mu_{V}~<~31~\rm{mag}~\rm{arcsec^{-2}}$ within $r_{\rm{200c}}$.

While this panel indicates the changes in ICL fraction by choice of faint-end surface brightness limit, it also reveals that the ICL fraction increases when focusing further inside due to the decreasing contribution of surviving satellites to the total stellar luminosity within the clusters.
However, the presence of the BCG stars significantly reduces the ICL fraction in the central region.
This highlights, for accurate comparisons between studies on ICL fraction, the need to carefully consider the effects of detection limits related to the observable radius and the faint-end surface brightness limit.

As the detection limits can influence the ICL fraction of the clusters, they can also change the relative contributions of different ICL formation channels.
The middle and bottom panels of Figure \ref{fig:profile} illustrate the cumulative ICL contribution as a function of distance from the BCG, normalized by $r_{\rm{200c}}$, depending on progenitor masses (the middle panel) and formation mechanisms (the bottom panel), using the surface brightness limit of $\mu_{V}=31~\rm{mag}~\rm{arcsec^{-2}}$.
In the middle panel, it is evident that the ``intermediate mass’’ satellites ($10^{10} < M_{\rm{gal}} [\msun] < 10^{11}$) consistently contribute as the main formation channel for ICL stars within $r_{\rm{200c}}$ of the clusters. 
Their contribution to the ICL increases when focusing on the inner region.
Conversely, the ``group mass" satellites ($M_{\rm{gal}} [\msun] > 10^{11}$) maintain a constant contribution to the ICL irrespective of the distance from the BCG.

The bottom panel reveals that disrupted ICL dominates near the BCG, where the strong tidal field more efficiently disrupts satellites. 
Conversely, preprocessed and stripped ICL stars contribute more as the distance from the BCG increases.
In particular, the stripping process dominates the formation of ICL stars within $r_{\rm{200c}}$ of the clusters, as shown in the bottom panels of Figure \ref{fig:ficl_type}.
Overall, these results show that the relative contributions of ICL formation channels depend on the detection limit, particularly the observable radius determined by the telescope’s field of view, and that even the dominant channel can change.
However, we note that the overall trends remain unchanged when adopting a brighter surface brightness limit of $\mu_{V}=28~\mathrm{mag\,arcsec^{-2}}$.

Moreover, \cite{tang2018} showed that the current measurement of ICL from observations and simulations also has a large dispersion due to differences in ICL definitions and observational environments (see Table 1 in their paper).
Indeed, \cite{chun2022} demonstrated that clear offsets ($\sim$5--20\%) in the ICL fraction of the same cluster can be observed between different ICL definitions.
These studies emphasize the importance of using the same method and observational parameters, such as surface brightness limit, PSF, and pixel size, to define the ICL and measure the ICL fraction.
As noted in Section \ref{sec:frac}, we will measure the ICL in more than a thousand clusters with a single instrument, and interpret these observations through the GRT simulations under the same observational limits.
In this respect, the synergy between GRT and K-DRIFT is expected to be highly effective.

\section{GRT-Driven science cases for K-DRIFT} \label{sec:case}
By enabling the detailed modeling of LSB structures across various environments, GRT provides a theoretical basis for interpreting the deep imaging data of K-DRIFT. 
In this section, we discuss key science cases where GRT simulations can be directly compared with K-DRIFT observations, illustrating the synergy between simulation and observation in understanding the formation and evolution of galaxies, groups, and clusters.

\begin{itemize}
\item {Diffuse light:

Analogous to the IGL and ICL within galaxy groups and clusters, the hierarchical merging of galaxies also produces diffuse light, such as stellar halos.
Such diffuse light provides valuable insights into the formation of galaxies \citep{contini2024b}.

Due to the deep photometric depth and wide FoV of K-DRIFT, we expect to observe the diffuse light of many galaxies, groups, and clusters within $\sim100~$Mpc volume.
This survey opens up the possibility of characterizing the diffuse light through measurements of luminosity fraction, radial profiles, and color gradients.
To interpret these observations, we have performed GRT simulations of more than 300 massive galaxies ($10^{12} < M_{\rm{200c}} < 5\times10^{12}$) and several tens of groups, as well as 84 clusters. 
Moreover, we prepare to perform additional GRT simulations targeting the structures in different environments.

While the GRT simulations have successively explained the gravitational evolution of the diffuse light without in-situ star formation within the clusters in previous works, it is important to consider that the contribution from the in-situ stars increases significantly in lower-mass structures \citep{pillepich2018}.
For this, we are currently developing a post-processing scheme that estimates the in-situ stellar component from the merger history of the structures. 
It enables us to investigate LSB structures on a galaxy scale.
Together with the upcoming K-DRIFT observations, the suite of GRT simulations will provide a detailed and statistically meaningful comparison between theory and observation, improving our understanding of the formation of structures.
}

\item {Tidal-featured galaxies:

The galaxies have undergone several merging events with accreted satellites or massive companions.
These merging events can leave distinct tidal features, such as tidal stream, tail, and shell-like structures.
The galaxies even form the tidal feature by themselves as they are gravitationally disrupted within a more massive structure.
Many previous observational and simulation studies have shown that the different morphology of tidal features may originate from distinct formation mechanisms, reflecting variations in orbital and merger histories, properties of progenitors, and the surrounding environment \citep[e.g.,][]{duc2015,mancillas2019,pop2018,yoon2024}.

As the suite of GRT simulations can trace the evolution of galaxies across a wide range of environments (e.g., fields, groups, and clusters), it enables us to connect the observed morphology of tidal features—such as streams, tails, and shells—with the merger history of each galaxy. 
This will open the possibility of interpreting the diverse tidal-featured galaxies that will be revealed by K-DRIFT, allowing us to statistically infer their evolutionary paths and the role of the environment in shaping their LSB features.
}

\item {Ultra-diffuse galaxies:

Ultra-diffuse galaxies (UDGs) have large physical sizes, comparable to those of massive galaxies, but exhibit extremely low surface brightness.
Following the discovery of numerous UDGs in the Coma cluster by \cite{vdokkum2015}, many studies have observed UDGs in various environments, including field, group, and cluster environments \citep[e.g.,][]{koda2015,mihos2015,vdokkum2015,iodice2020,lim2020,marleau2021}.
As UDGs exhibit extremely low surface brightness and a wide diversity of properties across different environments, they provide valuable constraints on how internal processes and environmental effects shape galaxy formation, especially in the low-mass regime.

There are two main formation scenarios for the UDGs: one scenario suggests that they are failed massive galaxies \citep{peng2016,toloba2018}, while the other proposes that they are puffed up by internal feedback or environmental effects \citep{dicinto2017,ogiya2018,toloba2018}.
In dense environments, such as galaxy groups and clusters, in particular, normal satellites can evolve into UDGs as they undergo tidal stripping and heating \citep[e.g.,][]{ogiya2018,sales2020,tremmel2020}.
These UDGs can have tidal features, and a recent observational study has indeed shown the presence of UDGs with tidal features \citep{zemaitis2023}.

Using the GRT simulations, \cite{chun2026} identified a total of 977 UDGs, with the number of UDGs per group or cluster ranging from 0 to 40. 
In their work, UDGs are defined as low-mass galaxies with $M_{\rm gal} < 10^{9}\,M_{\odot}$, low central surface brightness ($\mu_{g,0} > 24~\rm{mag~arcsec^{-2}}$), and large half-light radii ($R_{\rm e} > 1.5~\rm{kpc}$).
They found only two UDGs show their tidal features in the brighter surface brightness limit of $28~\rm{mag}~\rm{arcsec}^{-2}$ by visual inspection, thereby the tidal features of most UDGs are usually undiscovered in current observations.
On the other hand, deeper imaging observations with $\mu_{V}~<~31~\rm{mag}~\rm{arcsec}^{-2}$ increase the fraction of UDGs exhibiting tidal features, reaching about 14\% of the total UDGs.
This indicates that the upcoming deeper imaging observation with K-DRIFT will reveal evidence of tidal stripping of UDGs, such as the tidal tail near F8D1 UDG in the M81 group \citep{zemaitis2023}.
While the GRT does not model UDGs formed through hydrodynamical processes, it effectively traces the evolution of UDGs formed via gravitational tidal interactions in different environments (e.g., fields, groups, and clusters). 
By comparing the simulation results with K-DRIFT observations, it will be possible to identify tidal-origin UDGs and place constraints on their formation channels across various environments.

Moreover, once UDG-like objects are identified from the K-DRIFT observations, follow-up observations for the globular clusters (GCs) associated with their host UDGs are worthwhile, as they can provide valuable constraints on the total masses of UDGs. Although the accurate enclosed mass of a UDG can be derived from the kinematics of its stellar or gas components, it requires high-resolution spectroscopy with an extremely long exposure time due to its low surface brightness \citep{ems18,van19,dan20}.

Meanwhile, GCs associated with UDGs are relatively bright and spatially extended, and thus much easier to observe. GC kinematics can therefore be used to estimate the masses of UDGs \citep{van18,toloba2018,van19}. GC number counts can also be used for estimating its halo mass, as the total mass or number of GCs is known to correlate with the host halo mass in a wide mass range \citep{spi09,har17,van17,amo18,lim18,for20}.
Therefore, the statistically large UDG sample identified by K-DRIFT will provide ideal targets for subsequent GC observations, enabling constraints on the halo masses of UDGs and their environmental dependence.
To support such studies, we plan to incorporate the particle tagging method \citep[PTM;][Park et al. in prep.]{Park+2022} into the GRT simulations to model the coevolution of GC systems with their host UDGs in a cosmological context.
This will allow us to make theoretical predictions for GC populations associated with tidally formed UDGs and to quantify systematic uncertainties in halo mass estimates based on GC tracers.
}

\item {DM models:

DM constitutes approximately 27\% of the total energy density of the universe and about 85\% of the mass in galaxy clusters, making it an essential component in understanding the cosmos. However, its true nature remains elusive. The standard CDM model, despite its success on large scales, encounters several challenges on smaller scales, including the mismatch in the number of satellite galaxies (the ``missing satellite problem’’), their masses (the ``too-big-to-fail problem’’), and the central density profiles of galaxies (the ``core–cusp problem’’). These issues have motivated the development of alternative DM models, such as the self-interacting dark matter (SIDM) and fuzzy dark matter (FDM) scenarios.

SIDM introduces weak elastic interactions between DM particles, a feature naturally predicted in some particle physics frameworks \citep{2000PhRvL..84.3760S, 2000ApJ...534L.143B, 2018PhR...730....1T}.
It also offers potential solutions to the small-scale discrepancies seen in CDM-only simulations by reducing central densities and altering subhalo populations \citep{2017ARA&A..55..343B}.
Simulations with SIDM predict enhanced disruption of satellite galaxies and more extensive mass removal in MW-like galaxies and clusters compared to their CDM counterparts \citep[e.g.,][]{dooley2016, 2022MNRAS.511.5927S}.

FDM represents another compelling alternative, in which DM is composed of ultra-light axion-like particles with masses around $10^{-22}$ eV. The resulting de Broglie wavelength spans kpc scales, leading to quantum interference effects that suppress small-scale structure formation and provide a natural resolution to the core–cusp problem \citep{2000PhRvL..85.1158H, 2017PhRvD..95d3541H}. Simulations of FDM universes indicate that filaments appear visually smoother, reflecting the more coherent and less clumpy nature of DM accretion onto halos in FDM cosmologies. \citep{2019MNRAS.488.5551N, 2022JCAP...01..020S, 2023MNRAS.525..348D}. 

In conclusion, the differences among the DM models influence the survival and disruption of substructures within galaxies, groups, and clusters and thus directly affect the LSB features, including diffuse light and tidal features.
Since these features follow the global potential of the host structure and serve as visible probes of the underlying mass distribution \citep{yoo2022, 2024ApJ...965..145Y, 2025ApJ...988..229Y}, comparing their spatial distributions with those of DM in simulations with different DM models provides a way to constrain the nature of DM.
However, implementing SIDM with a range of cross-sections or FDM with a variety of particle masses—and especially doing so with the inclusion of baryonic physics—demands substantial computational resources. This is one of the main limitations of current SIDM and FDM simulations, which are constrained in terms of box size, resolution, sample size of structures, and the diversity of physical settings such as cross-sections or DM masses. Our GRT simulations, serving as an alternative to full cosmological hydrodynamical simulations, are expected to provide an efficient and effective means of overcoming these limitations.
While GRT simulations do not fully capture the impact of baryonic physics on LSB features, they provide an efficient framework for comparative studies across different DM models, for example by examining how they affect the formation and survivability of UDGs in clusters \citep[e.g.,][]{ragagnin2024}, the spatial distribution of the ICL (e.g., Yoo et al. in prep.), and the morphology of tidal features \citep[e.g.,][]{hainje2025}.
These simulations can then be compared with deep K-DRIFT observations, providing constraints on the nature of DM through visible tracers.
}

\item {
Comparison with cosmological hydrodynamic simulations:

The GRT is designed to efficiently describe the tidal-driven dynamical evolution of stellar components in a hierarchical context, but due to a lack of hydrodynamics, it has a severe limitation in that ongoing/in-situ star formation is not included. More specifically, the limitation is prominent in those LSB studies: (1) ram pressure stripping of satellite galaxies and the subsequent star formation activities, (2) the in-situ star formation of UDGs, which highly originate from hydrodynamical processes, and (3) the color gradient across the LSB features, which may reflect past merging and both in-situ and ex-situ star formation histories. To overcome the limitation, we aim to mimic the in-situ star formation through post-processing as mentioned above. However, this approach still cannot capture the self-consistency of hydrodynamic simulations. Therefore, it is necessary to compare the GRT results with hydrodynamic simulations conducted using high-resolution elements that describe the LSB regime. 

For this, we will adopt three different cosmological hydrodynamic simulations: Horizon Run 5 \citep{lee21}, NewCluster \citep{han2026}, and DARWIN (in prep). The Horizon Run 5 covers the large-scale universe on the Gpc scale with a resolution of 1~kpc, which is beneficial for statistical studies, particularly those for galaxy clusters. The main limitation of the Horizon Run 5 is that the calculation halts at redshift 0.625. Although the sample is limited to a single galaxy cluster, the NewCluster has an advantage in a much higher resolution of 70~pc, which is still ongoing beyond redshift 0.7. It can provide valuable statistics on the tidal and ram-pressure stripping of satellite galaxies formed in a dense, cluster environment. Since the NewCluster captures both the in-situ and tidal origin of the UDG formation, it can be valuable for studying the nature of the UDGs in the cluster environment. The LSB features in various environments can be accessed through the DARWIN, which consists of dozens of zoom-in simulations targeting dwarf galaxies in a wide range of environments--from void-like low-density regions to high-density galaxy groups--with a high resolution of 62.5~pc. Each feature of these simulations is believed to complement the limitations of the GRT.}
\end{itemize}

\section{Summary}\label{sec:summary}
The K-DRIFT project aims to explore LSB structures in the nearby universe, which provide valuable clues to the formation and evolution of structures, such as galaxies, groups, and clusters.
To interpret the observation data from K-DRIFT, the simulation study is essential.
As K-DRIFT is designed to detect such diffuse features in many structures with a surface brightness reaching $\sim$30 mag arcsec$^{-2}$, the simulations must not only resolve the stellar components with high spatial and mass resolution, but also provide a statistically large sample to enable meaningful comparisons.
For this, the GRT has been developed.

The GRT provides an efficient and physically motivated framework for modeling LSB structures in a cosmological context.
As such, it is particularly well-suited to support the science goals of K-DRIFT.
The GRT simulations can generate realistic mock surface brightness maps that reach below $30~\rm{mag}~\rm{arcsec}^{-2}$, enabling direct comparisons with the detection limits of K-DRIFT.
These simulations can trace the evolution of properties and spatial distribution of LSB features, thereby establishing a theoretical basis for interpreting K-DRIFT observations.
Additionally, the computational efficiency of GRT allows for the construction of statistically significant samples across a wide range of environments.
Recent studies have applied this method to 84 galaxy clusters, confirming its suitability for LSB studies \citep[][]{yoo2022,chun2023,chun2024,chun2026}.
This makes it possible to connect observed LSB structures to their merger histories.
Finally, because the GRT traces the dynamical evolution of stellar components from their origin to their present configuration, it provides a powerful framework for inferring the physical mechanisms of observed LSB features.
By linking the morphology of LSB features to merger histories and environmental effects, GRT simulations can help disentangle the roles of mergers and stripping in shaping the faint structures within galaxies, groups, and clusters.
Therefore, the GRT serves as a crucial theoretical basis for interpreting the observations of K-DRIFT.
Together, K-DRIFT and GRT will represent a powerful synergy between observation and theory, offering a unique opportunity for a deeper understanding of the formation and evolution of galaxies and the large-scale structures of the universe.


\acknowledgments




 \bibliography{main}

@ARTICLE{lee21,
       author = {{Lee}, Jaehyun and {Shin}, Jihye and {Snaith}, Owain N. and {Kim}, Yonghwi and {Few}, C. Gareth and {Devriendt}, Julien and {Dubois}, Yohan and {Cox}, Leah M. and {Hong}, Sungwook E. and {Kwon}, Oh-Kyoung and {Park}, Chan and {Pichon}, Christophe and {Kim}, Juhan and {Gibson}, Brad K. and {Park}, Changbom},
        title = "{The Horizon Run 5 Cosmological Hydrodynamical Simulation: Probing Galaxy Formation from Kilo- to Gigaparsec Scales}",
      journal = {\apj},
         year = 2021,
        month = feb,
       volume = {908},
       number = {1},
          eid = {11},
        pages = {11},
          doi = {10.3847/1538-4357/abd08b},
archivePrefix = {arXiv},
       eprint = {2006.01039},
 primaryClass = {astro-ph.GA},
       adsurl = {https://ui.adsabs.harvard.edu/abs/2021ApJ...908...11L}
}

@ARTICLE{2022JCAP...01..020S,
       author = {{Sabiu}, Cristiano G. and {Kadota}, Kenji and {Asorey}, Jacobo and {Park}, Inkyu},
        title = "{Probing ultra-light axion dark matter from 21 cm tomography using Convolutional Neural Networks}",
      journal = {\jcap},
         year = 2022,
        month = jan,
       volume = {2022},
       number = {1},
          eid = {020},
        pages = {020},
          doi = {10.1088/1475-7516/2022/01/020},
archivePrefix = {arXiv},
       eprint = {2108.07972},
 primaryClass = {astro-ph.CO},
       adsurl = {https://ui.adsabs.harvard.edu/abs/2022JCAP...01..020S}
}

@ARTICLE{2019MNRAS.488.5551N,
       author = {{Ni}, Yueying and {Wang}, Mei-Yu and {Feng}, Yu and {Di Matteo}, Tiziana},
        title = "{Predictions for the abundance of high-redshift galaxies in a fuzzy dark matter universe}",
      journal = {\mnras},
         year = 2019,
        month = oct,
       volume = {488},
       number = {4},
        pages = {5551-5565},
          doi = {10.1093/mnras/stz2085},
archivePrefix = {arXiv},
       eprint = {1904.01604},
 primaryClass = {astro-ph.CO},
       adsurl = {https://ui.adsabs.harvard.edu/abs/2019MNRAS.488.5551N}
}

@ARTICLE{2023MNRAS.525..348D,
       author = {{Dome}, Tibor and {Fialkov}, Anastasia and {Sartorio}, Nina and {Mocz}, Philip},
        title = "{Cosmic web dissection in fuzzy dark matter cosmologies}",
      journal = {\mnras},
         year = 2023,
        month = oct,
       volume = {525},
       number = {1},
        pages = {348-363},
          doi = {10.1093/mnras/stad2276},
archivePrefix = {arXiv},
       eprint = {2301.09762},
 primaryClass = {astro-ph.CO},
       adsurl = {https://ui.adsabs.harvard.edu/abs/2023MNRAS.525..348D}
}

@ARTICLE{2017PhRvD..95d3541H,
       author = {{Hui}, Lam and {Ostriker}, Jeremiah P. and {Tremaine}, Scott and {Witten}, Edward},
        title = "{Ultralight scalars as cosmological dark matter}",
      journal = {\prd},
         year = 2017,
        month = feb,
       volume = {95},
       number = {4},
          eid = {043541},
        pages = {043541},
          doi = {10.1103/PhysRevD.95.043541},
archivePrefix = {arXiv},
       eprint = {1610.08297},
 primaryClass = {astro-ph.CO},
       adsurl = {https://ui.adsabs.harvard.edu/abs/2017PhRvD..95d3541H}
}

@ARTICLE{2000PhRvL..85.1158H,
       author = {{Hu}, Wayne and {Barkana}, Rennan and {Gruzinov}, Andrei},
        title = "{Fuzzy Cold Dark Matter: The Wave Properties of Ultralight Particles}",
      journal = {\prl},
         year = 2000,
        month = aug,
       volume = {85},
       number = {6},
        pages = {1158-1161},
          doi = {10.1103/PhysRevLett.85.1158},
archivePrefix = {arXiv},
       eprint = {astro-ph/0003365},
 primaryClass = {astro-ph},
       adsurl = {https://ui.adsabs.harvard.edu/abs/2000PhRvL..85.1158H}
}

@ARTICLE{2000PhRvL..84.3760S,
       author = {{Spergel}, David N. and {Steinhardt}, Paul J.},
        title = "{Observational Evidence for Self-Interacting Cold Dark Matter}",
      journal = {\prl},
         year = 2000,
        month = apr,
       volume = {84},
       number = {17},
        pages = {3760-3763},
          doi = {10.1103/PhysRevLett.84.3760},
archivePrefix = {arXiv},
       eprint = {astro-ph/9909386},
 primaryClass = {astro-ph},
       adsurl = {https://ui.adsabs.harvard.edu/abs/2000PhRvL..84.3760S}
}

@ARTICLE{2017ARA&A..55..343B,
       author = {{Bullock}, James S. and {Boylan-Kolchin}, Michael},
        title = "{Small-Scale Challenges to the {\ensuremath{\Lambda}}CDM Paradigm}",
      journal = {\araa},
         year = 2017,
        month = aug,
       volume = {55},
       number = {1},
        pages = {343-387},
          doi = {10.1146/annurev-astro-091916-055313},
archivePrefix = {arXiv},
       eprint = {1707.04256},
 primaryClass = {astro-ph.CO},
       adsurl = {https://ui.adsabs.harvard.edu/abs/2017ARA&A..55..343B}
}

@ARTICLE{2022MNRAS.511.5927S,
       author = {{Sirks}, Ellen L. and {Oman}, Kyle A. and {Robertson}, Andrew and {Massey}, Richard and {Frenk}, Carlos},
        title = "{The effects of self-interacting dark matter on the stripping of galaxies that fall into clusters}",
      journal = {\mnras},
         year = 2022,
        month = apr,
       volume = {511},
       number = {4},
        pages = {5927-5935},
          doi = {10.1093/mnras/stac406},
archivePrefix = {arXiv},
       eprint = {2109.03257},
 primaryClass = {astro-ph.CO},
       adsurl = {https://ui.adsabs.harvard.edu/abs/2022MNRAS.511.5927S}
}

@ARTICLE{2000ApJ...534L.143B,
       author = {{Burkert}, Andreas},
        title = "{The Structure and Evolution of Weakly Self-interacting Cold Dark Matter Halos}",
      journal = {\apjl},
         year = 2000,
        month = may,
       volume = {534},
       number = {2},
        pages = {L143-L146},
          doi = {10.1086/312674},
archivePrefix = {arXiv},
       eprint = {astro-ph/0002409},
 primaryClass = {astro-ph},
       adsurl = {https://ui.adsabs.harvard.edu/abs/2000ApJ...534L.143B}
}

@ARTICLE{2018PhR...730....1T,
       author = {{Tulin}, Sean and {Yu}, Hai-Bo},
        title = "{Dark matter self-interactions and small scale structure}",
      journal = {\physrep},
         year = 2018,
        month = feb,
       volume = {730},
        pages = {1-57},
          doi = {10.1016/j.physrep.2017.11.004},
archivePrefix = {arXiv},
       eprint = {1705.02358},
 primaryClass = {hep-ph},
       adsurl = {https://ui.adsabs.harvard.edu/abs/2018PhR...730....1T}
}

@ARTICLE{2024ApJ...965..145Y,
       author = {{Yoo}, Jaewon and {Park}, Changbom and {Sabiu}, Cristiano G. and {Singh}, Ankit and {Ko}, Jongwan and {Lee}, Jaehyun and {Pichon}, Christophe and {Jee}, M. James and {Gibson}, Brad K. and {Snaith}, Owain and {Kim}, Juhan and {Shin}, Jihye and {Kim}, Yonghwi and {Kim}, Hyowon},
        title = "{Spatial Distribution of Intracluster Light versus Dark Matter in Horizon Run 5}",
      journal = {\apj},
         year = 2024,
        month = apr,
       volume = {965},
       number = {2},
          eid = {145},
        pages = {145},
          doi = {10.3847/1538-4357/ad2df8},
archivePrefix = {arXiv},
       eprint = {2402.17958},
 primaryClass = {astro-ph.CO},
       adsurl = {https://ui.adsabs.harvard.edu/abs/2024ApJ...965..145Y}
}

@ARTICLE{2025ApJ...988..229Y,
       author = {{Yoo}, Jaewon and {Shin}, Jihye and {Hwang}, Ho Seong and {Sabiu}, Cristiano G. and {Kim}, Hyowon and {Ko}, Jongwan and {Lee}, Jong Chul},
        title = "{Tracing Dark Matter in the Central Regions of Galaxy Clusters Using Galaxies, Gas, and Intracluster Light in TNG300: Connections to Cluster Dynamical State}",
      journal = {\apj},
         year = 2025,
        month = aug,
       volume = {988},
       number = {2},
          eid = {229},
        pages = {229},
          doi = {10.3847/1538-4357/ade66f},
archivePrefix = {arXiv},
       eprint = {2506.16280},
 primaryClass = {astro-ph.GA},
       adsurl = {https://ui.adsabs.harvard.edu/abs/2025ApJ...988..229Y}
}

@ARTICLE{har17,
       author = {{Harris}, William E. and {Blakeslee}, John P. and {Harris}, Gretchen L.~H.},
        title = "{Galactic Dark Matter Halos and Globular Cluster Populations. III. Extension to Extreme Environments}",
      journal = {\apj},
         year = 2017,
        month = feb,
       volume = {836},
       number = {1},
          eid = {67},
        pages = {67},
          doi = {10.3847/1538-4357/836/1/67},
archivePrefix = {arXiv},
       eprint = {1701.04845},
 primaryClass = {astro-ph.GA},
       adsurl = {https://ui.adsabs.harvard.edu/abs/2017ApJ...836...67H}
}

@ARTICLE{for20,
       author = {{Forbes}, Duncan A. and {Alabi}, Adebusola and {Romanowsky}, Aaron J. and {Brodie}, Jean P. and {Arimoto}, Nobuo},
        title = "{Globular clusters in Coma cluster ultra-diffuse galaxies (UDGs): evidence for two types of UDG?}",
      journal = {\mnras},
         year = 2020,
        month = mar,
       volume = {492},
       number = {4},
        pages = {4874-4883},
          doi = {10.1093/mnras/staa180},
archivePrefix = {arXiv},
       eprint = {2001.10031},
 primaryClass = {astro-ph.GA},
       adsurl = {https://ui.adsabs.harvard.edu/abs/2020MNRAS.492.4874F}
}

@ARTICLE{spi09,
       author = {{Spitler}, L.~R. and {Forbes}, D.~A.},
        title = "{A new method for estimating dark matter halo masses using globular cluster systems}",
      journal = {\mnras},
         year = 2009,
        month = jan,
       volume = {392},
       number = {1},
        pages = {L1-L5},
          doi = {10.1111/j.1745-3933.2008.00567.x},
archivePrefix = {arXiv},
       eprint = {0809.5057},
 primaryClass = {astro-ph},
       adsurl = {https://ui.adsabs.harvard.edu/abs/2009MNRAS.392L...1S}
}

@ARTICLE{lim18,
       author = {{Lim}, Sungsoon and {Peng}, Eric W. and {C{\^o}t{\'e}}, Patrick and {Sales}, Laura V. and {den Brok}, Mark and {Blakeslee}, John P. and {Guhathakurta}, Puragra},
        title = "{The Globular Cluster Systems of Ultra-diffuse Galaxies in the Coma Cluster}",
      journal = {\apj},
         year = 2018,
        month = jul,
       volume = {862},
       number = {1},
          eid = {82},
        pages = {82},
          doi = {10.3847/1538-4357/aacb81},
archivePrefix = {arXiv},
       eprint = {1806.05425},
 primaryClass = {astro-ph.GA},
       adsurl = {https://ui.adsabs.harvard.edu/abs/2018ApJ...862...82L}
}

@ARTICLE{amo18,
       author = {{Amorisco}, N.~C. and {Monachesi}, A. and {Agnello}, A. and {White}, S.~D.~M.},
        title = "{The globular cluster systems of 54 Coma ultra-diffuse galaxies: statistical constraints from HST data}",
      journal = {\mnras},
         year = 2018,
        month = apr,
       volume = {475},
       number = {3},
        pages = {4235-4251},
          doi = {10.1093/mnras/sty116},
archivePrefix = {arXiv},
       eprint = {1610.01595},
 primaryClass = {astro-ph.GA},
       adsurl = {https://ui.adsabs.harvard.edu/abs/2018MNRAS.475.4235A}
}

@ARTICLE{van17,
       author = {{van Dokkum}, Pieter and {Abraham}, Roberto and {Romanowsky}, Aaron J. and {Brodie}, Jean and {Conroy}, Charlie and {Danieli}, Shany and {Lokhorst}, Deborah and {Merritt}, Allison and {Mowla}, Lamiya and {Zhang}, Jielai},
        title = "{Extensive Globular Cluster Systems Associated with Ultra Diffuse Galaxies in the Coma Cluster}",
      journal = {\apjl},
         year = 2017,
        month = jul,
       volume = {844},
       number = {1},
          eid = {L11},
        pages = {L11},
          doi = {10.3847/2041-8213/aa7ca2},
archivePrefix = {arXiv},
       eprint = {1705.08513},
 primaryClass = {astro-ph.GA},
       adsurl = {https://ui.adsabs.harvard.edu/abs/2017ApJ...844L..11V}
}

@ARTICLE{dan20,
       author = {{Danieli}, Shany and {van Dokkum}, Pieter and {Abraham}, Roberto and {Conroy}, Charlie and {Dolphin}, Andrew E. and {Romanowsky}, Aaron J.},
        title = "{A Tip of the Red Giant Branch Distance to the Dark Matter Deficient Galaxy NGC 1052-DF4 from Deep Hubble Space Telescope Data}",
      journal = {\apjl},
         year = 2020,
        month = may,
       volume = {895},
       number = {1},
          eid = {L4},
        pages = {L4},
          doi = {10.3847/2041-8213/ab8dc4},
archivePrefix = {arXiv},
       eprint = {1910.07529},
 primaryClass = {astro-ph.GA},
       adsurl = {https://ui.adsabs.harvard.edu/abs/2020ApJ...895L...4D}
}

@ARTICLE{van19,
       author = {{van Dokkum}, Pieter and {Danieli}, Shany and {Abraham}, Roberto and {Conroy}, Charlie and {Romanowsky}, Aaron J.},
        title = "{A Second Galaxy Missing Dark Matter in the NGC 1052 Group}",
      journal = {\apjl},
         year = 2019,
        month = mar,
       volume = {874},
       number = {1},
          eid = {L5},
        pages = {L5},
          doi = {10.3847/2041-8213/ab0d92},
archivePrefix = {arXiv},
       eprint = {1901.05973},
 primaryClass = {astro-ph.GA},
       adsurl = {https://ui.adsabs.harvard.edu/abs/2019ApJ...874L...5V}
}

@ARTICLE{ems18,
       author = {{Emsellem}, Eric and {van der Burg}, Remco F.~J. and {Fensch}, J{\'e}r{\'e}my and {Je{\v{r}}{\'a}bkov{\'a}}, Tereza and {Zanella}, Anita and {Agnello}, Adriano and {Hilker}, Michael and {M{\"u}ller}, Oliver and {Rejkuba}, Marina and {Duc}, Pierre-Alain and {Durrell}, Patrick and {Habas}, Rebecca and {Lelli}, Federico and {Lim}, Sungsoon and {Marleau}, Francine R. and {Peng}, Eric and {S{\'a}nchez-Janssen}, Rub{\'e}n},
        title = "{The ultra-diffuse galaxy NGC 1052-DF2 with MUSE. I. Kinematics of the stellar body}",
      journal = {\aap},
         year = 2019,
        month = may,
       volume = {625},
          eid = {A76},
        pages = {A76},
          doi = {10.1051/0004-6361/201834909},
archivePrefix = {arXiv},
       eprint = {1812.07345},
 primaryClass = {astro-ph.GA},
       adsurl = {https://ui.adsabs.harvard.edu/abs/2019A&A...625A..76E}
}

@inproceedings{chang2025,
	author = {Seunghyuk Chang},
	booktitle = {UV/Optical/IR Space Telescopes and Instruments: Innovative Technologies and Concepts VI},
	doi = {10.1117/12.2023433},
	editor = {Howard A. MacEwen and James B. Breckinridge},
	organization = {International Society for Optics and Photonics},
	pages = {219 -- 228},
	publisher = {SPIE},
	title = {{Elimination of linear astigmatism in off-axis three-mirror telescope and its applications}},
	url = {https://doi.org/10.1117/12.2023433},
	volume = {8860},
	year = {2013}}

@ARTICLE{van18,
       author = {{van Dokkum}, Pieter and {Danieli}, Shany and {Cohen}, Yotam and {Merritt}, Allison and {Romanowsky}, Aaron J. and {Abraham}, Roberto and {Brodie}, Jean and {Conroy}, Charlie and {Lokhorst}, Deborah and {Mowla}, Lamiya and {O'Sullivan}, Ewan and {Zhang}, Jielai},
        title = "{A galaxy lacking dark matter}",
      journal = {\nat},
         year = 2018,
        month = mar,
       volume = {555},
       number = {7698},
        pages = {629-632},
          doi = {10.1038/nature25767},
archivePrefix = {arXiv},
       eprint = {1803.10237},
 primaryClass = {astro-ph.GA},
       adsurl = {https://ui.adsabs.harvard.edu/abs/2018Natur.555..629V}
}

@ARTICLE{agertz2011,
       author = {{Agertz}, Oscar and {Teyssier}, Romain and {Moore}, Ben},
        title = "{The formation of disc galaxies in a {\ensuremath{\Lambda}}CDM universe}",
      journal = {\mnras},
         year = 2011,
        month = jan,
       volume = {410},
       number = {2},
        pages = {1391-1408},
          doi = {10.1111/j.1365-2966.2010.17530.x},
archivePrefix = {arXiv},
       eprint = {1004.0005},
 primaryClass = {astro-ph.CO},
       adsurl = {https://ui.adsabs.harvard.edu/abs/2011MNRAS.410.1391A}
}

@ARTICLE{amorisco2015,
       author = {{Amorisco}, N.~C.},
        title = "{On feathers, bifurcations and shells: the dynamics of tidal streams across the mass scale}",
      journal = {\mnras},
         year = 2015,
        month = jun,
       volume = {450},
       number = {1},
        pages = {575-591},
          doi = {10.1093/mnras/stv648},
archivePrefix = {arXiv},
       eprint = {1410.0360},
 primaryClass = {astro-ph.GA},
       adsurl = {https://ui.adsabs.harvard.edu/abs/2015MNRAS.450..575A}
}

@ARTICLE{awad2023,
       author = {{Awad}, Petra and {Peletier}, Reynier and {Canducci}, Marco and {Smith}, Rory and {Taghribi}, Abolfazl and {Mohammadi}, Mohammad and {Shin}, Jihye and {Ti{\v{n}}o}, Peter and {Bunte}, Kerstin},
        title = "{Swarm-intelligence-based extraction and manifold crawling along the Large-Scale Structure}",
      journal = {\mnras},
         year = 2023,
        month = apr,
       volume = {520},
       number = {3},
        pages = {4517-4539},
          doi = {10.1093/mnras/stad428},
archivePrefix = {arXiv},
       eprint = {2302.03779},
 primaryClass = {astro-ph.IM},
       adsurl = {https://ui.adsabs.harvard.edu/abs/2023MNRAS.520.4517A}
}

@ARTICLE{behroozi2013a,
       author = {{Behroozi}, Peter S. and {Wechsler}, Risa H. and {Conroy}, Charlie},
        title = "{The Average Star Formation Histories of Galaxies in Dark Matter Halos from z = 0-8}",
      journal = {\apj},
         year = 2013,
        month = jun,
       volume = {770},
       number = {1},
          eid = {57},
        pages = {57},
          doi = {10.1088/0004-637X/770/1/57},
archivePrefix = {arXiv},
       eprint = {1207.6105},
 primaryClass = {astro-ph.CO},
       adsurl = {https://ui.adsabs.harvard.edu/abs/2013ApJ...770...57B}
}

@ARTICLE{behroozi2013b,
       author = {{Behroozi}, Peter S. and {Wechsler}, Risa H. and {Wu}, Hao-Yi},
        title = "{The ROCKSTAR Phase-space Temporal Halo Finder and the Velocity Offsets of Cluster Cores}",
      journal = {\apj},
         year = 2013,
        month = jan,
       volume = {762},
       number = {2},
          eid = {109},
        pages = {109},
          doi = {10.1088/0004-637X/762/2/109},
archivePrefix = {arXiv},
       eprint = {1110.4372},
 primaryClass = {astro-ph.CO},
       adsurl = {https://ui.adsabs.harvard.edu/abs/2013ApJ...762..109B}
}

@ARTICLE{behroozi2013c,
       author = {{Behroozi}, Peter S. and {Wechsler}, Risa H. and {Wu}, Hao-Yi and {Busha}, Michael T. and {Klypin}, Anatoly A. and {Primack}, Joel R.},
        title = "{Gravitationally Consistent Halo Catalogs and Merger Trees for Precision Cosmology}",
      journal = {\apj},
         year = 2013,
        month = jan,
       volume = {763},
       number = {1},
          eid = {18},
        pages = {18},
          doi = {10.1088/0004-637X/763/1/18},
archivePrefix = {arXiv},
       eprint = {1110.4370},
 primaryClass = {astro-ph.CO},
       adsurl = {https://ui.adsabs.harvard.edu/abs/2013ApJ...763...18B}
}

@ARTICLE{bruzual2003,
       author = {{Bruzual}, G. and {Charlot}, S.},
        title = "{Stellar population synthesis at the resolution of 2003}",
      journal = {\mnras},
         year = 2003,
        month = oct,
       volume = {344},
       number = {4},
        pages = {1000-1028},
          doi = {10.1046/j.1365-8711.2003.06897.x},
archivePrefix = {arXiv},
       eprint = {astro-ph/0309134},
 primaryClass = {astro-ph},
       adsurl = {https://ui.adsabs.harvard.edu/abs/2003MNRAS.344.1000B}
}

@ARTICLE{burke2015,
       author = {{Burke}, Claire and {Hilton}, Matt and {Collins}, Chris},
        title = "{Coevolution of brightest cluster galaxies and intracluster light using CLASH}",
      journal = {\mnras},
         year = 2015,
        month = may,
       volume = {449},
       number = {3},
        pages = {2353-2367},
          doi = {10.1093/mnras/stv450},
archivePrefix = {arXiv},
       eprint = {1503.04321},
 primaryClass = {astro-ph.CO},
       adsurl = {https://ui.adsabs.harvard.edu/abs/2015MNRAS.449.2353B}
}

@ARTICLE{byun2022,
       author = {{Byun}, Woowon and {Ko}, Jongwan and {Kim}, Yunjong and {Seon}, Kwang-Il and {Chang}, Seunghyuk and {Kim}, Dohoon and {Choi}, Changsu and {Chun}, Sang-Hyun and {Jeon}, Young-Beom and {Kim}, Jae-Woo and {Lee}, Chung-Uk and {Lee}, Yongseok and {Park}, Hong Soo and {Sung}, Eon-Chang and {Yoo}, Jaewon and {Lee}, Gayoung and {Lee}, Hyoungkwon},
        title = "{Performance Assessment of the KASI-Deep Rolling Imaging Fast-optics Telescope Pathfinder}",
      journal = {\pasp},
         year = 2022,
        month = aug,
       volume = {134},
       number = {1038},
          eid = {084101},
        pages = {084101},
          doi = {10.1088/1538-3873/ac8500},
archivePrefix = {arXiv},
       eprint = {2207.14553},
 primaryClass = {astro-ph.GA},
       adsurl = {https://ui.adsabs.harvard.edu/abs/2022PASP..134h4101B}
}

@ARTICLE{byun2026,
       author = {{Byun}, Woowon and {Yoon}, Yongmin and {Ko}, Jongwan and {Lee}, Yun Hee and {Lee}, Gain and {Hwang}, Ho Seong and {Sabiu}, Cristiano G. and {Seon}, Kwang-il and {Chun}, Kyungwon and {Shin}, Jihye and {Rhee}, Jinsu and {Kim}, Jae-Woo and {Yoo}, Jaewon and {Lee}, Jaehyun and {Chun}, Sang-Hyun and {Park}, Hong Soo and {Yang}, Soung-Chul and {Hong}, Sungryong and {Shin}, Jeehye and {Kim}, Hyowon},
        title = "{K-DRIFT Science Theme: Galaxies in the Faint Universe}",
      journal = {arXiv e-prints},
         year = 2026,
        month = feb,
          eid = {arXiv:2602.08283},
        pages = {arXiv:2602.08283},
archivePrefix = {arXiv},
       eprint = {2602.08283},
 primaryClass = {astro-ph.GA},
       adsurl = {https://ui.adsabs.harvard.edu/abs/2026arXiv260208283B}
}

@ARTICLE{chun2022,
       author = {{Chun}, Kyungwon and {Shin}, Jihye and {Smith}, Rory and {Ko}, Jongwan and {Yoo}, Jaewon},
        title = "{The Galaxy Replacement Technique (GRT): A New Approach to Study Tidal Stripping and Formation of Intracluster Light in a Cosmological Context}",
      journal = {\apj},
         year = 2022,
        month = feb,
       volume = {925},
       number = {2},
          eid = {103},
        pages = {103},
          doi = {10.3847/1538-4357/ac2cbe},
archivePrefix = {arXiv},
       eprint = {2110.09532},
 primaryClass = {astro-ph.GA},
       adsurl = {https://ui.adsabs.harvard.edu/abs/2022ApJ...925..103C}
}

@ARTICLE{chun2023,
       author = {{Chun}, Kyungwon and {Shin}, Jihye and {Smith}, Rory and {Ko}, Jongwan and {Yoo}, Jaewon},
        title = "{The Formation of the Brightest Cluster Galaxy and Intracluster Light in Cosmological N-body Simulations with the Galaxy Replacement Technique}",
      journal = {\apj},
         year = 2023,
        month = feb,
       volume = {943},
       number = {2},
          eid = {148},
        pages = {148},
          doi = {10.3847/1538-4357/aca890},
archivePrefix = {arXiv},
       eprint = {2212.02510},
 primaryClass = {astro-ph.GA},
       adsurl = {https://ui.adsabs.harvard.edu/abs/2023ApJ...943..148C}
}

@ARTICLE{chun2024,
       author = {{Chun}, Kyungwon and {Shin}, Jihye and {Ko}, Jongwan and {Smith}, Rory and {Yoo}, Jaewon},
        title = "{Formation Channels of Diffuse Lights in Groups and Clusters over Time}",
      journal = {\apj},
         year = 2024,
        month = jul,
       volume = {969},
       number = {2},
          eid = {142},
        pages = {142},
          doi = {10.3847/1538-4357/ad4a52},
archivePrefix = {arXiv},
       eprint = {2405.08061},
 primaryClass = {astro-ph.GA},
       adsurl = {https://ui.adsabs.harvard.edu/abs/2024ApJ...969..142C}
}

@ARTICLE{chun2026,
       author = {{Chun}, Kyungwon and {Shin}, Jihye and {Ko}, Jongwan and {Smith}, Rory and {Park}, So-Myoung and {Nam}, Songhee},
        title = "{The Role of Preprocessing in Tidal Feature Formation within Galaxy Clusters}",
      journal = {\apj},
         year = 2026,
        month = jan,
       volume = {996},
       number = {2},
          eid = {145},
        pages = {145},
          doi = {10.3847/1538-4357/ae0cc7},
archivePrefix = {arXiv},
       eprint = {2509.22802},
 primaryClass = {astro-ph.GA},
       adsurl = {https://ui.adsabs.harvard.edu/abs/2026ApJ...996..145C}
}

@ARTICLE{conroy2007,
       author = {{Conroy}, Charlie and {Wechsler}, Risa H. and {Kravtsov}, Andrey V.},
        title = "{The Hierarchical Build-Up of Massive Galaxies and the Intracluster Light since z = 1}",
      journal = {\apj},
         year = 2007,
        month = oct,
       volume = {668},
       number = {2},
        pages = {826-838},
          doi = {10.1086/521425},
archivePrefix = {arXiv},
       eprint = {astro-ph/0703374},
 primaryClass = {astro-ph},
       adsurl = {https://ui.adsabs.harvard.edu/abs/2007ApJ...668..826C}
}

@ARTICLE{contini2018,
       author = {{Contini}, E. and {Yi}, S.~K. and {Kang}, X.},
        title = "{The different growth pathways of brightest cluster galaxies and intracluster light}",
      journal = {\mnras},
         year = 2018,
        month = sep,
       volume = {479},
       number = {1},
        pages = {932-944},
          doi = {10.1093/mnras/sty1518},
archivePrefix = {arXiv},
       eprint = {1806.01480},
 primaryClass = {astro-ph.GA},
       adsurl = {https://ui.adsabs.harvard.edu/abs/2018MNRAS.479..932C}
}

@ARTICLE{contini2019,
       author = {{Contini}, E. and {Yi}, S.~K. and {Kang}, X.},
        title = "{Theoretical Predictions of Colors and Metallicity of the Intracluster Light}",
      journal = {\apj},
         year = 2019,
        month = jan,
       volume = {871},
       number = {1},
          eid = {24},
        pages = {24},
          doi = {10.3847/1538-4357/aaf41f},
archivePrefix = {arXiv},
       eprint = {1811.03253},
 primaryClass = {astro-ph.GA},
       adsurl = {https://ui.adsabs.harvard.edu/abs/2019ApJ...871...24C}
}

@ARTICLE{contini2021,
       author = {{Contini}, Emanuele},
        title = "{On the Origin and Evolution of the Intra-Cluster Light: A Brief Review of the Most Recent Developments}",
      journal = {Galaxies},
         year = 2021,
        month = aug,
       volume = {9},
       number = {3},
        pages = {60},
          doi = {10.3390/galaxies9030060},
archivePrefix = {arXiv},
       eprint = {2107.04180},
 primaryClass = {astro-ph.GA},
       adsurl = {https://ui.adsabs.harvard.edu/abs/2021Galax...9...60C}
}

@ARTICLE{contini2023,
       author = {{Contini}, Emanuele and {Jeon}, Seyoung and {Rhee}, Jinsu and {Han}, San and {Yi}, Sukyoung K.},
        title = "{The Intracluster Light and Its Link with the Dynamical State of the Host Group/Cluster: The Role of the Halo Concentration}",
      journal = {\apj},
         year = 2023,
        month = nov,
       volume = {958},
       number = {1},
          eid = {72},
        pages = {72},
          doi = {10.3847/1538-4357/acfd25},
archivePrefix = {arXiv},
       eprint = {2310.03263},
 primaryClass = {astro-ph.GA},
       adsurl = {https://ui.adsabs.harvard.edu/abs/2023ApJ...958...72C}
}

@ARTICLE{contini2024,
       author = {{Contini}, Emanuele and {Rhee}, Jinsu and {Han}, San and {Jeon}, Seyoung and {Yi}, Sukyoung K.},
        title = "{The Connection between the Intracluster Light and its Host Halo: Formation Time and Contribution from Different Channels}",
      journal = {\aj},
         year = 2024,
        month = jan,
       volume = {167},
       number = {1},
          eid = {7},
        pages = {7},
          doi = {10.3847/1538-3881/ad0894},
archivePrefix = {arXiv},
       eprint = {2310.20135},
 primaryClass = {astro-ph.GA},
       adsurl = {https://ui.adsabs.harvard.edu/abs/2024AJ....167....7C}
}

@ARTICLE{contini2024b,
       author = {{Contini}, Emanuele and {Han}, San and {Jeon}, Seyoung and {Rhee}, Jinsu and {Yi}, Sukyoung K.},
        title = "{Diffuse Light in Milky Way{\textendash}like Haloes}",
      journal = {\apjl},
         year = 2024,
        month = feb,
       volume = {962},
       number = {1},
          eid = {L10},
        pages = {L10},
          doi = {10.3847/2041-8213/ad21e2},
archivePrefix = {arXiv},
       eprint = {2401.14650},
 primaryClass = {astro-ph.GA},
       adsurl = {https://ui.adsabs.harvard.edu/abs/2024ApJ...962L..10C}
}

@ARTICLE{contreras2024,
       author = {{Contreras-Santos}, A. and {Knebe}, A. and {Cui}, W. and {Alonso Asensio}, I. and {Dalla Vecchia}, C. and {Ca{\~n}as}, R. and {Haggar}, R. and {Mostoghiu Paun}, R.~A. and {Pearce}, F.~R. and {Rasia}, E.},
        title = "{Characterising the intra-cluster light in The Three Hundred simulations}",
      journal = {\aap},
         year = 2024,
        month = mar,
       volume = {683},
          eid = {A59},
        pages = {A59},
          doi = {10.1051/0004-6361/202348474},
archivePrefix = {arXiv},
       eprint = {2401.08283},
 primaryClass = {astro-ph.CO},
       adsurl = {https://ui.adsabs.harvard.edu/abs/2024A&A...683A..59C}
}

@ARTICLE{cooper2010,
       author = {{Cooper}, A.~P. and {Cole}, S. and {Frenk}, C.~S. and {White}, S.~D.~M. and {Helly}, J. and {Benson}, A.~J. and {De Lucia}, G. and {Helmi}, A. and {Jenkins}, A. and {Navarro}, J.~F. and {Springel}, V. and {Wang}, J.},
        title = "{Galactic stellar haloes in the CDM model}",
      journal = {\mnras},
         year = 2010,
        month = aug,
       volume = {406},
       number = {2},
        pages = {744-766},
          doi = {10.1111/j.1365-2966.2010.16740.x},
archivePrefix = {arXiv},
       eprint = {0910.3211},
 primaryClass = {astro-ph.GA},
       adsurl = {https://ui.adsabs.harvard.edu/abs/2010MNRAS.406..744C}
}

@ARTICLE{cooper2011,
       author = {{Cooper}, Andrew P. and {Mart{\'\i}nez-Delgado}, David and {Helly}, John and {Frenk}, Carlos and {Cole}, Shaun and {Crawford}, Ken and {Zibetti}, Stefano and {Carballo-Bello}, Julio A. and {GaBany}, R. Jay},
        title = "{The Formation of Shell Galaxies Similar to NGC 7600 in the Cold Dark Matter Cosmogony}",
      journal = {\apjl},
         year = 2011,
        month = dec,
       volume = {743},
       number = {1},
          eid = {L21},
        pages = {L21},
          doi = {10.1088/2041-8205/743/1/L21},
archivePrefix = {arXiv},
       eprint = {1111.2864},
 primaryClass = {astro-ph.GA},
       adsurl = {https://ui.adsabs.harvard.edu/abs/2011ApJ...743L..21C}
}

@ARTICLE{darocha2008,
       author = {{Da Rocha}, C. and {Ziegler}, B.~L. and {Mendes de Oliveira}, C.},
        title = "{Intragroup diffuse light in compact groups of galaxies - II. HCG 15, 35 and 51}",
      journal = {\mnras},
         year = 2008,
        month = aug,
       volume = {388},
       number = {3},
        pages = {1433-1443},
          doi = {10.1111/j.1365-2966.2008.13500.x},
archivePrefix = {arXiv},
       eprint = {0805.4015},
 primaryClass = {astro-ph},
       adsurl = {https://ui.adsabs.harvard.edu/abs/2008MNRAS.388.1433D}
}

@ARTICLE{dicinto2017,
       author = {{Di Cintio}, Arianna and {Brook}, Chris B. and {Dutton}, Aaron A. and {Macci{\`o}}, Andrea V. and {Obreja}, Aura and {Dekel}, Avishai},
        title = "{NIHAO - XI. Formation of ultra-diffuse galaxies by outflows}",
      journal = {\mnras},
         year = 2017,
        month = mar,
       volume = {466},
       number = {1},
        pages = {L1-L6},
          doi = {10.1093/mnrasl/slw210},
archivePrefix = {arXiv},
       eprint = {1608.01327},
 primaryClass = {astro-ph.GA},
       adsurl = {https://ui.adsabs.harvard.edu/abs/2017MNRAS.466L...1D}
}

@ARTICLE{dong2024,
       author = {{Dong}, Kyung Lin and {Smith}, Rory and {Shin}, Jihye and {Peletier}, Reynier},
        title = "{Enhanced destruction of cluster satellites by major mergers}",
      journal = {\mnras},
         year = 2024,
        month = jan,
       volume = {527},
       number = {3},
        pages = {9185-9191},
          doi = {10.1093/mnras/stad3799},
archivePrefix = {arXiv},
       eprint = {2312.04641},
 primaryClass = {astro-ph.GA},
       adsurl = {https://ui.adsabs.harvard.edu/abs/2024MNRAS.527.9185D}
}

@ARTICLE{dooley2016,
       author = {{Dooley}, Gregory A. and {Peter}, Annika H.~G. and {Vogelsberger}, Mark and {Zavala}, Jes{\'u}s and {Frebel}, Anna},
        title = "{Enhanced tidal stripping of satellites in the galactic halo from dark matter self-interactions}",
      journal = {\mnras},
         year = 2016,
        month = sep,
       volume = {461},
       number = {1},
        pages = {710-727},
          doi = {10.1093/mnras/stw1309},
archivePrefix = {arXiv},
       eprint = {1603.08919},
 primaryClass = {astro-ph.GA},
       adsurl = {https://ui.adsabs.harvard.edu/abs/2016MNRAS.461..710D}
}

@ARTICLE{dubois2021,
       author = {{Dubois}, Yohan and {Beckmann}, Ricarda and {Bournaud}, Fr{\'e}d{\'e}ric and {Choi}, Hoseung and {Devriendt}, Julien and {Jackson}, Ryan and {Kaviraj}, Sugata and {Kimm}, Taysun and {Kraljic}, Katarina and {Laigle}, Clotilde and {Martin}, Garreth and {Park}, Min-Jung and {Peirani}, S{\'e}bastien and {Pichon}, Christophe and {Volonteri}, Marta and {Yi}, Sukyoung K.},
        title = "{Introducing the NEWHORIZON simulation: Galaxy properties with resolved internal dynamics across cosmic time}",
      journal = {\aap},
         year = 2021,
        month = jul,
       volume = {651},
          eid = {A109},
        pages = {A109},
          doi = {10.1051/0004-6361/202039429},
archivePrefix = {arXiv},
       eprint = {2009.10578},
 primaryClass = {astro-ph.GA},
       adsurl = {https://ui.adsabs.harvard.edu/abs/2021A&A...651A.109D}
}

@ARTICLE{duc2015,
       author = {{Duc}, Pierre-Alain and {Cuillandre}, Jean-Charles and {Karabal}, Emin and {Cappellari}, Michele and {Alatalo}, Katherine and {Blitz}, Leo and {Bournaud}, Fr{\'e}d{\'e}ric and {Bureau}, Martin and {Crocker}, Alison F. and {Davies}, Roger L. and {Davis}, Timothy A. and {de Zeeuw}, P.~T. and {Emsellem}, Eric and {Khochfar}, Sadegh and {Krajnovi{\'c}}, Davor and {Kuntschner}, Harald and {McDermid}, Richard M. and {Michel-Dansac}, Leo and {Morganti}, Raffaella and {Naab}, Thorsten and {Oosterloo}, Tom and {Paudel}, Sanjaya and {Sarzi}, Marc and {Scott}, Nicholas and {Serra}, Paolo and {Weijmans}, Anne-Marie and {Young}, Lisa M.},
        title = "{The ATLAS$^{3D}$ project - XXIX. The new look of early-type galaxies and surrounding fields disclosed by extremely deep optical images}",
      journal = {\mnras},
         year = 2015,
        month = jan,
       volume = {446},
       number = {1},
        pages = {120-143},
          doi = {10.1093/mnras/stu2019},
archivePrefix = {arXiv},
       eprint = {1410.0981},
 primaryClass = {astro-ph.GA},
       adsurl = {https://ui.adsabs.harvard.edu/abs/2015MNRAS.446..120D}
}

@ARTICLE{dutton2009,
       author = {{Dutton}, Aaron A.},
        title = "{On the origin of exponential galaxy discs}",
      journal = {\mnras},
         year = 2009,
        month = jun,
       volume = {396},
       number = {1},
        pages = {121-140},
          doi = {10.1111/j.1365-2966.2009.14741.x},
archivePrefix = {arXiv},
       eprint = {0810.5164},
 primaryClass = {astro-ph},
       adsurl = {https://ui.adsabs.harvard.edu/abs/2009MNRAS.396..121D}
}

@ARTICLE{dutton2011,
       author = {{Dutton}, Aaron A. and {van den Bosch}, Frank C. and {Faber}, Sandra M. and {Simard}, Luc and {Kassin}, Susan A. and {Koo}, David C. and {Bundy}, Kevin and {Huang}, Jiasheng and {Weiner}, Benjamin J. and {Cooper}, Michael C. and {Newman}, Jeffrey A. and {Mozena}, Mark and {Koekemoer}, Anton M.},
        title = "{On the evolution of the velocity-mass-size relations of disc-dominated galaxies over the past 10 billion years}",
      journal = {\mnras},
         year = 2011,
        month = jan,
       volume = {410},
       number = {3},
        pages = {1660-1676},
          doi = {10.1111/j.1365-2966.2010.17555.x},
archivePrefix = {arXiv},
       eprint = {1006.3558},
 primaryClass = {astro-ph.GA},
       adsurl = {https://ui.adsabs.harvard.edu/abs/2011MNRAS.410.1660D}
}

@ARTICLE{fall1980,
       author = {{Fall}, S.~M. and {Efstathiou}, G.},
        title = "{Formation and rotation of disc galaxies with haloes.}",
      journal = {\mnras},
         year = 1980,
        month = oct,
       volume = {193},
        pages = {189-206},
          doi = {10.1093/mnras/193.2.189},
       adsurl = {https://ui.adsabs.harvard.edu/abs/1980MNRAS.193..189F}
}

@ARTICLE{furnell2021,
       author = {{Furnell}, Kate E. and {Collins}, Chris A. and {Kelvin}, Lee S. and {Baldry}, Ivan K. and {James}, Phil A. and {Manolopoulou}, Maria and {Mann}, Robert G. and {Giles}, Paul A. and {Bermeo}, Alberto and {Hilton}, Matthew and {Wilkinson}, Reese and {Romer}, A. Kathy and {Vergara}, Carlos and {Bhargava}, Sunayana and {Stott}, John P. and {Mayers}, Julian and {Viana}, Pedro},
        title = "{The growth of intracluster light in XCS-HSC galaxy clusters from 0.1 < z < 0.5}",
      journal = {\mnras},
         year = 2021,
        month = apr,
       volume = {502},
       number = {2},
        pages = {2419-2437},
          doi = {10.1093/mnras/stab065},
archivePrefix = {arXiv},
       eprint = {2101.01644},
 primaryClass = {astro-ph.GA},
       adsurl = {https://ui.adsabs.harvard.edu/abs/2021MNRAS.502.2419F}
}

@ARTICLE{genel2018,
       author = {{Genel}, Shy and {Nelson}, Dylan and {Pillepich}, Annalisa and {Springel}, Volker and {Pakmor}, R{\"u}diger and {Weinberger}, Rainer and {Hernquist}, Lars and {Naiman}, Jill and {Vogelsberger}, Mark and {Marinacci}, Federico and {Torrey}, Paul},
        title = "{The size evolution of star-forming and quenched galaxies in the IllustrisTNG simulation}",
      journal = {\mnras},
         year = 2018,
        month = mar,
       volume = {474},
       number = {3},
        pages = {3976-3996},
          doi = {10.1093/mnras/stx3078},
archivePrefix = {arXiv},
       eprint = {1707.05327},
 primaryClass = {astro-ph.GA},
       adsurl = {https://ui.adsabs.harvard.edu/abs/2018MNRAS.474.3976G}
}

@ARTICLE{haggar2020,
       author = {{Haggar}, Roan and {Gray}, Meghan E. and {Pearce}, Frazer R. and {Knebe}, Alexander and {Cui}, Weiguang and {Mostoghiu}, Robert and {Yepes}, Gustavo},
        title = "{The Three Hundred project: backsplash galaxies in simulations of clusters}",
      journal = {\mnras},
         year = 2020,
        month = mar,
       volume = {492},
       number = {4},
        pages = {6074-6085},
          doi = {10.1093/mnras/staa273},
archivePrefix = {arXiv},
       eprint = {2001.11518},
 primaryClass = {astro-ph.GA},
       adsurl = {https://ui.adsabs.harvard.edu/abs/2020MNRAS.492.6074H}
}

@ARTICLE{Hahn2011,
       author = {{Hahn}, Oliver and {Abel}, Tom},
        title = "{Multi-scale initial conditions for cosmological simulations}",
      journal = {\mnras},
         year = 2011,
        month = aug,
       volume = {415},
       number = {3},
        pages = {2101-2121},
          doi = {10.1111/j.1365-2966.2011.18820.x},
archivePrefix = {arXiv},
       eprint = {1103.6031},
 primaryClass = {astro-ph.CO},
       adsurl = {https://ui.adsabs.harvard.edu/abs/2011MNRAS.415.2101H}
}

@ARTICLE{hainje2025,
       author = {{Hainje}, Connor and {Slone}, Oren and {Lisanti}, Mariangela and {Erkal}, Denis},
        title = "{Effects of Dark Matter Self-interactions on Sagittarius and its Stream}",
      journal = {\apj},
         year = 2025,
        month = nov,
       volume = {993},
       number = {1},
          eid = {6},
        pages = {6},
          doi = {10.3847/1538-4357/adfed8},
archivePrefix = {arXiv},
       eprint = {2503.15589},
 primaryClass = {astro-ph.GA},
       adsurl = {https://ui.adsabs.harvard.edu/abs/2025ApJ...993....6H}
}

@ARTICLE{han2026,
       author = {{Han}, S. and {Yi}, S.~K. and {Dubois}, Y. and {Rhee}, J. and {Jeon}, S. and {Jang}, J.~K. and {Byun}, G.-H. and {Cadiou}, C. and {Kim}, J. and {Kimm}, T. and {Pichon}, C.},
        title = "{Introducing NewCluster: First half of the history of a high-resolution cluster simulation}",
      journal = {\aap},
         year = 2026,
        month = jan,
       volume = {705},
          eid = {A169},
        pages = {A169},
          doi = {10.1051/0004-6361/202556291},
archivePrefix = {arXiv},
       eprint = {2507.06301},
 primaryClass = {astro-ph.GA},
       adsurl = {https://ui.adsabs.harvard.edu/abs/2026A&A...705A.169H}
}

@ARTICLE{iodice2020,
       author = {{Iodice}, E. and {Cantiello}, M. and {Hilker}, M. and {Rejkuba}, M. and {Arnaboldi}, M. and {Spavone}, M. and {Greggio}, L. and {Forbes}, D.~A. and {D'Ago}, G. and {Mieske}, S. and {Spiniello}, C. and {La Marca}, A. and {Rampazzo}, R. and {Paolillo}, M. and {Capaccioli}, M. and {Schipani}, P.},
        title = "{The first detection of ultra-diffuse galaxies in the Hydra I cluster from the VEGAS survey}",
      journal = {\aap},
         year = 2020,
        month = oct,
       volume = {642},
          eid = {A48},
        pages = {A48},
          doi = {10.1051/0004-6361/202038523},
archivePrefix = {arXiv},
       eprint = {2007.11533},
 primaryClass = {astro-ph.GA},
       adsurl = {https://ui.adsabs.harvard.edu/abs/2020A&A...642A..48I}
}

@ARTICLE{jeon2026,
       author = {{Jeon}, Seyoung and {Contini}, Emanuele and {Han}, San and {Rhee}, Jinsu and {Martin}, Garreth and {Kim}, Juhan and {Lee}, Jaehyun and {Kimm}, Taysun and {Pichon}, Christophe and {Byun}, Gyeong-Hwan and {Dubois}, Yohan and {Cadiou}, Corentin and {Jang}, J.~K. and {Yi}, Sukyoung K.},
        title = "{On the Origin of Intracluster Light Based on the High-resolution Simulation, NEWCLUSTER}",
      journal = {\apj},
         year = 2026,
        month = feb,
       volume = {998},
       number = {1},
          eid = {30},
        pages = {30},
          doi = {10.3847/1538-4357/ae2eaa},
archivePrefix = {arXiv},
       eprint = {2512.06098},
 primaryClass = {astro-ph.GA},
       adsurl = {https://ui.adsabs.harvard.edu/abs/2026ApJ...998...30J}
}

@ARTICLE{jhee2022,
       author = {{Jhee}, Hannah and {Song}, Hyunmi and {Smith}, Rory and {Shin}, Jihye and {Park}, Inkyu and {Laigle}, Clotilde},
        title = "{Tracking Halo Orbits and Their Mass Evolution around Large-scale Filaments}",
      journal = {\apj},
         year = 2022,
        month = nov,
       volume = {940},
       number = {1},
          eid = {2},
        pages = {2},
          doi = {10.3847/1538-4357/ac990a},
archivePrefix = {arXiv},
       eprint = {2201.09540},
 primaryClass = {astro-ph.CO},
       adsurl = {https://ui.adsabs.harvard.edu/abs/2022ApJ...940....2J}
}

@ARTICLE{ji2014,
       author = {{Ji}, Inchan and {Peirani}, S{\'e}bastien and {Yi}, Sukyoung K.},
        title = "{Lifetime of merger features of equal-mass disk mergers}",
      journal = {\aap},
         year = 2014,
        month = jun,
       volume = {566},
          eid = {A97},
        pages = {A97},
          doi = {10.1051/0004-6361/201423530},
archivePrefix = {arXiv},
       eprint = {1405.1807},
 primaryClass = {astro-ph.GA},
       adsurl = {https://ui.adsabs.harvard.edu/abs/2014A&A...566A..97J}
}

@ARTICLE{jimenez-teja2018,
       author = {{Jim{\'e}nez-Teja}, Yolanda and {Dupke}, Renato and {Ben{\'\i}tez}, Narciso and {Koekemoer}, Anton M. and {Zitrin}, Adi and {Umetsu}, Keiichi and {Ziegler}, Bodo L. and {Frye}, Brenda L. and {Ford}, Holland and {Bouwens}, Rychard J. and {Bradley}, Larry D. and {Broadhurst}, Thomas and {Coe}, Dan and {Donahue}, Megan and {Graves}, Genevieve J. and {Grillo}, Claudio and {Infante}, Leopoldo and {Jouvel}, Stephanie and {Kelson}, Daniel D. and {Lahav}, Ofer and {Lazkoz}, Ruth and {Lemze}, Dorom and {Maoz}, Dan and {Medezinski}, Elinor and {Melchior}, Peter and {Meneghetti}, Massimo and {Mercurio}, Amata and {Merten}, Julian and {Molino}, Alberto and {Moustakas}, Leonidas A. and {Nonino}, Mario and {Ogaz}, Sara and {Riess}, Adam G. and {Rosati}, Piero and {Sayers}, Jack and {Seitz}, Stella and {Zheng}, Wei},
        title = "{Unveiling the Dynamical State of Massive Clusters through the ICL Fraction}",
      journal = {\apj},
         year = 2018,
        month = apr,
       volume = {857},
       number = {2},
          eid = {79},
        pages = {79},
          doi = {10.3847/1538-4357/aab70f},
archivePrefix = {arXiv},
       eprint = {1803.04981},
 primaryClass = {astro-ph.GA},
       adsurl = {https://ui.adsabs.harvard.edu/abs/2018ApJ...857...79J}
}

@ARTICLE{jimenez-teja2024,
       author = {{Jim{\'e}nez-Teja}, Yolanda and {Dupke}, Renato A. and {Lopes}, Paulo A.~A. and {Dimauro}, Paola},
        title = "{Evidence for a Redshifted Excess in the Intracluster Light Fractions of Merging Clusters at z   0.8}",
      journal = {\apjl},
         year = 2024,
        month = jan,
       volume = {960},
       number = {2},
          eid = {L7},
        pages = {L7},
          doi = {10.3847/2041-8213/ad181a},
archivePrefix = {arXiv},
       eprint = {2401.02543},
 primaryClass = {astro-ph.GA},
       adsurl = {https://ui.adsabs.harvard.edu/abs/2024ApJ...960L...7J}
}

@ARTICLE{joo2023,
       author = {{Joo}, Hyungjin and {Jee}, M. James},
        title = "{Intracluster light is already abundant at redshift beyond unity}",
      journal = {\nat},
         year = 2023,
        month = jan,
       volume = {613},
       number = {7942},
        pages = {37-41},
          doi = {10.1038/s41586-022-05396-4},
archivePrefix = {arXiv},
       eprint = {2301.01523},
 primaryClass = {astro-ph.GA},
       adsurl = {https://ui.adsabs.harvard.edu/abs/2023Natur.613...37J}
}

@ARTICLE{joo2025,
       author = {{Joo}, Hyungjin and {Jee}, M. James and {Kim}, Juhan and {Lee}, Jaehyun and {Ko}, Jongwan and {Park}, Changbom and {Shin}, Jihye and {Snaith}, Owain and {Pichon}, Christophe and {Gibson}, Brad and {Kim}, Yonghwi},
        title = "{Tracing the Formation History of Intrahalo Light with Horizon Run 5}",
      journal = {\apj},
         year = 2025,
        month = sep,
       volume = {990},
       number = {2},
          eid = {96},
        pages = {96},
          doi = {10.3847/1538-4357/adf4d0},
archivePrefix = {arXiv},
       eprint = {2411.08117},
 primaryClass = {astro-ph.GA},
       adsurl = {https://ui.adsabs.harvard.edu/abs/2025ApJ...990...96J}
}

@ARTICLE{khalid2024,
       author = {{Khalid}, A. and {Brough}, S. and {Martin}, G. and {Kimmig}, L.~C. and {Lagos}, C.~D.~P. and {Remus}, R. -S. and {Martinez-Lombilla}, C.},
        title = "{Characterizing tidal features around galaxies in cosmological simulations}",
      journal = {\mnras},
         year = 2024,
        month = jun,
       volume = {530},
       number = {4},
        pages = {4422-4445},
          doi = {10.1093/mnras/stae1064},
archivePrefix = {arXiv},
       eprint = {2404.12436},
 primaryClass = {astro-ph.GA},
       adsurl = {https://ui.adsabs.harvard.edu/abs/2024MNRAS.530.4422K}
}

@ARTICLE{kim2022,
       author = {{Kim}, Yigon and {Smith}, Rory and {Shin}, Jihye},
        title = "{Unexpected Dancing Partners: Tracing the Coherence between the Spin and Motion of Dark Matter Halos}",
      journal = {\apj},
         year = 2022,
        month = aug,
       volume = {935},
       number = {2},
          eid = {71},
        pages = {71},
          doi = {10.3847/1538-4357/ac7e45},
archivePrefix = {arXiv},
       eprint = {2207.02389},
 primaryClass = {astro-ph.CO},
       adsurl = {https://ui.adsabs.harvard.edu/abs/2022ApJ...935...71K}
}

@ARTICLE{kim2024,
       author = {{Kim}, Hyowon and {Smith}, Rory and {Ko}, Jongwan and {Shinn}, Jong-Ho and {Chun}, Kyungwon and {Shin}, Jihye and {Yoo}, Jaewon},
        title = "{New Observational Recipes for Measuring Dynamical States of Galaxy Clusters}",
      journal = {\apj},
         year = 2024,
        month = aug,
       volume = {970},
       number = {2},
          eid = {165},
        pages = {165},
          doi = {10.3847/1538-4357/ad4f80},
archivePrefix = {arXiv},
       eprint = {2405.06245},
 primaryClass = {astro-ph.GA},
       adsurl = {https://ui.adsabs.harvard.edu/abs/2024ApJ...970..165K}
}

@ARTICLE{ko2018,
       author = {{Ko}, Jongwan and {Jee}, M. James},
        title = "{Evidence for the Existence of Abundant Intracluster Light at z = 1.24}",
      journal = {\apj},
         year = 2018,
        month = aug,
       volume = {862},
       number = {2},
          eid = {95},
        pages = {95},
          doi = {10.3847/1538-4357/aacbda},
archivePrefix = {arXiv},
       eprint = {1806.02687},
 primaryClass = {astro-ph.GA},
       adsurl = {https://ui.adsabs.harvard.edu/abs/2018ApJ...862...95K}
}

@ARTICLE{ko2025,
       author = {{Ko}, Jongwan and {Byun}, Woowon and {Seon}, Kwang-Il and {Kim}, Jihun and {Kim}, Yunjong and {Kim}, Daewook and {Chang}, Seunghyuk and {Kim}, Dohoon and {Kweon Moon}, Il and {Kwon}, Hyuksun and {Kim}, Yeonsik and {Ahn}, Kyohoon and {Lee}, Gayoung and {Lee}, Yongseok and {Lee}, Sangmin and {Cha}, Sang-Mok and {Kim}, Dong-Jin and {Park}, Kyusu and {Yoo}, Jaewon and {Kim}, Jae-Woo and {Shin}, Jihye and {Chun}, Sang-Hyun and {Yoon}, Yongmin and {Lee}, Jaehyun and {Chun}, Kyungwon and {Rhee}, Jinsu and {Hong}, Sungryong and {Park}, Jongyeob and {Jeon}, Young-Beom and {Sung}, Eon-Chang and {Park}, Hong Soo and {Kim}, Seonwoo and {Bahk}, GyeongGon and {Yeon}, Seri},
        title = "{K-DRIFT: Unveiling New Imagery of the Hidden Universe}",
      journal = {arXiv e-prints},
         year = 2025,
        month = oct,
          eid = {arXiv:2510.22250},
        pages = {arXiv:2510.22250},
archivePrefix = {arXiv},
       eprint = {2510.22250},
 primaryClass = {astro-ph.GA},
       adsurl = {https://ui.adsabs.harvard.edu/abs/2025arXiv251022250K}
}

@ARTICLE{koda2015,
       author = {{Koda}, Jin and {Yagi}, Masafumi and {Yamanoi}, Hitomi and {Komiyama}, Yutaka},
        title = "{Approximately a Thousand Ultra-diffuse Galaxies in the Coma Cluster}",
      journal = {\apjl},
         year = 2015,
        month = jul,
       volume = {807},
       number = {1},
          eid = {L2},
        pages = {L2},
          doi = {10.1088/2041-8205/807/1/L2},
archivePrefix = {arXiv},
       eprint = {1506.01712},
 primaryClass = {astro-ph.GA},
       adsurl = {https://ui.adsabs.harvard.edu/abs/2015ApJ...807L...2K}
}

@ARTICLE{lim2020,
       author = {{Lim}, Sungsoon and {C{\^o}t{\'e}}, Patrick and {Peng}, Eric W. and {Ferrarese}, Laura and {Roediger}, Joel C. and {Durrell}, Patrick R. and {Mihos}, J. Christopher and {Wang}, Kaixiang and {Gwyn}, S.~D.~J. and {Cuillandre}, Jean-Charles and {Liu}, Chengze and {S{\'a}nchez-Janssen}, Rub{\'e}n and {Toloba}, Elisa and {Sales}, Laura V. and {Guhathakurta}, Puragra and {Lan{\c{c}}on}, Ariane and {Puzia}, Thomas H.},
        title = "{The Next Generation Virgo Cluster Survey (NGVS). XXX. Ultra-diffuse Galaxies and Their Globular Cluster Systems}",
      journal = {\apj},
         year = 2020,
        month = aug,
       volume = {899},
       number = {1},
          eid = {69},
        pages = {69},
          doi = {10.3847/1538-4357/aba433},
archivePrefix = {arXiv},
       eprint = {2007.10565},
 primaryClass = {astro-ph.GA},
       adsurl = {https://ui.adsabs.harvard.edu/abs/2020ApJ...899...69L}
}

@ARTICLE{ma2016,
       author = {{Ma}, Xiangcheng and {Hopkins}, Philip F. and {Faucher-Gigu{\`e}re}, Claude-Andr{\'e} and {Zolman}, Nick and {Muratov}, Alexander L. and {Kere{\v{s}}}, Du{\v{s}}an and {Quataert}, Eliot},
        title = "{The origin and evolution of the galaxy mass-metallicity relation}",
      journal = {\mnras},
         year = 2016,
        month = feb,
       volume = {456},
       number = {2},
        pages = {2140-2156},
          doi = {10.1093/mnras/stv2659},
archivePrefix = {arXiv},
       eprint = {1504.02097},
 primaryClass = {astro-ph.GA},
       adsurl = {https://ui.adsabs.harvard.edu/abs/2016MNRAS.456.2140M}
}

@ARTICLE{mancillas2019,
       author = {{Mancillas}, Brisa and {Duc}, Pierre-Alain and {Combes}, Fran{\c{c}}oise and {Bournaud}, Fr{\'e}d{\'e}ric and {Emsellem}, Eric and {Martig}, Marie and {Michel-Dansac}, Leo},
        title = "{Probing the merger history of red early-type galaxies with their faint stellar substructures}",
      journal = {\aap},
         year = 2019,
        month = dec,
       volume = {632},
          eid = {A122},
        pages = {A122},
          doi = {10.1051/0004-6361/201936320},
archivePrefix = {arXiv},
       eprint = {1909.07500},
 primaryClass = {astro-ph.GA},
       adsurl = {https://ui.adsabs.harvard.edu/abs/2019A&A...632A.122M}
}

@ARTICLE{marleau2021,
       author = {{Marleau}, Francine R. and {Habas}, Rebecca and {Poulain}, M{\'e}lina and {Duc}, Pierre-Alain and {M{\"u}ller}, Oliver and {Lim}, Sungsoon and {Durrell}, Patrick R. and {S{\'a}nchez-Janssen}, Rub{\'e}n and {Paudel}, Sanjaya and {Ahad}, Syeda Lammim and {Chougule}, Abhishek and {B{\'\i}lek}, Michal and {Fensch}, J{\'e}r{\'e}my},
        title = "{Ultra diffuse galaxies in the MATLAS low-to-moderate density fields}",
      journal = {\aap},
         year = 2021,
        month = oct,
       volume = {654},
          eid = {A105},
        pages = {A105},
          doi = {10.1051/0004-6361/202141432},
archivePrefix = {arXiv},
       eprint = {2109.13173},
 primaryClass = {astro-ph.GA},
       adsurl = {https://ui.adsabs.harvard.edu/abs/2021A&A...654A.105M}
}

@ARTICLE{martin2022,
       author = {{Martin}, G. and {Bazkiaei}, A.~E. and {Spavone}, M. and {Iodice}, E. and {Mihos}, J.~C. and {Montes}, M. and {Benavides}, J.~A. and {Brough}, S. and {Carlin}, J.~L. and {Collins}, C.~A. and {Duc}, P.~A. and {G{\'o}mez}, F.~A. and {Galaz}, G. and {Hern{\'a}ndez-Toledo}, H.~M. and {Jackson}, R.~A. and {Kaviraj}, S. and {Knapen}, J.~H. and {Mart{\'\i}nez-Lombilla}, C. and {McGee}, S. and {O'Ryan}, D. and {Prole}, D.~J. and {Rich}, R.~M. and {Rom{\'a}n}, J. and {Shah}, E.~A. and {Starkenburg}, T.~K. and {Watkins}, A.~E. and {Zaritsky}, D. and {Pichon}, C. and {Armus}, L. and {Bianconi}, M. and {Buitrago}, F. and {Bus{\'a}}, I. and {Davis}, F. and {Demarco}, R. and {Desmons}, A. and {Garc{\'\i}a}, P. and {Graham}, A.~W. and {Holwerda}, B. and {Hon}, D.~S. -H. and {Khalid}, A. and {Klehammer}, J. and {Klutse}, D.~Y. and {Lazar}, I. and {Nair}, P. and {Noakes-Kettel}, E.~A. and {Rutkowski}, M. and {Saha}, K. and {Sahu}, N. and {Sola}, E. and {V{\'a}zquez-Mata}, J.~A. and {Vera-Casanova}, A. and {Yoon}, I.},
        title = "{Preparing for low surface brightness science with the Vera C. Rubin Observatory: Characterization of tidal features from mock images}",
      journal = {\mnras},
         year = 2022,
        month = jun,
       volume = {513},
       number = {1},
        pages = {1459-1487},
          doi = {10.1093/mnras/stac1003},
archivePrefix = {arXiv},
       eprint = {2203.07675},
 primaryClass = {astro-ph.GA},
       adsurl = {https://ui.adsabs.harvard.edu/abs/2022MNRAS.513.1459M}
}

@ARTICLE{mihos2005,
       author = {{Mihos}, J. Christopher and {Harding}, Paul and {Feldmeier}, John and {Morrison}, Heather},
        title = "{Diffuse Light in the Virgo Cluster}",
      journal = {\apjl},
         year = 2005,
        month = sep,
       volume = {631},
       number = {1},
        pages = {L41-L44},
          doi = {10.1086/497030},
archivePrefix = {arXiv},
       eprint = {astro-ph/0508217},
 primaryClass = {astro-ph},
       adsurl = {https://ui.adsabs.harvard.edu/abs/2005ApJ...631L..41M}
}

@ARTICLE{mihos2015,
       author = {{Mihos}, J. Christopher and {Durrell}, Patrick R. and {Ferrarese}, Laura and {Feldmeier}, John J. and {C{\^o}t{\'e}}, Patrick and {Peng}, Eric W. and {Harding}, Paul and {Liu}, Chengze and {Gwyn}, Stephen and {Cuillandre}, Jean-Charles},
        title = "{Galaxies at the Extremes: Ultra-diffuse Galaxies in the Virgo Cluster}",
      journal = {\apjl},
         year = 2015,
        month = aug,
       volume = {809},
       number = {2},
          eid = {L21},
        pages = {L21},
          doi = {10.1088/2041-8205/809/2/L21},
archivePrefix = {arXiv},
       eprint = {1507.02270},
 primaryClass = {astro-ph.GA},
       adsurl = {https://ui.adsabs.harvard.edu/abs/2015ApJ...809L..21M}
}

@ARTICLE{mihos2017,
       author = {{Mihos}, J. Christopher and {Harding}, Paul and {Feldmeier}, John J. and {Rudick}, Craig and {Janowiecki}, Steven and {Morrison}, Heather and {Slater}, Colin and {Watkins}, Aaron},
        title = "{The Burrell Schmidt Deep Virgo Survey: Tidal Debris, Galaxy Halos, and Diffuse Intracluster Light in the Virgo Cluster}",
      journal = {\apj},
         year = 2017,
        month = jan,
       volume = {834},
       number = {1},
          eid = {16},
        pages = {16},
          doi = {10.3847/1538-4357/834/1/16},
archivePrefix = {arXiv},
       eprint = {1611.04435},
 primaryClass = {astro-ph.GA},
       adsurl = {https://ui.adsabs.harvard.edu/abs/2017ApJ...834...16M}
}

@ARTICLE{montes2014,
       author = {{Montes}, Mireia and {Trujillo}, Ignacio},
        title = "{Intracluster Light at the Frontier: A2744}",
      journal = {\apj},
         year = 2014,
        month = oct,
       volume = {794},
       number = {2},
          eid = {137},
        pages = {137},
          doi = {10.1088/0004-637X/794/2/137},
archivePrefix = {arXiv},
       eprint = {1405.2070},
 primaryClass = {astro-ph.CO},
       adsurl = {https://ui.adsabs.harvard.edu/abs/2014ApJ...794..137M}
}

@ARTICLE{montes2018,
       author = {{Montes}, Mireia and {Trujillo}, Ignacio},
        title = "{Intracluster light at the Frontier - II. The Frontier Fields Clusters}",
      journal = {\mnras},
         year = 2018,
        month = feb,
       volume = {474},
       number = {1},
        pages = {917-932},
          doi = {10.1093/mnras/stx2847},
archivePrefix = {arXiv},
       eprint = {1710.03240},
 primaryClass = {astro-ph.CO},
       adsurl = {https://ui.adsabs.harvard.edu/abs/2018MNRAS.474..917M}
}

@ARTICLE{montes2022,
       author = {{Montes}, Mireia},
        title = "{The faint light in groups and clusters of galaxies}",
      journal = {Nature Astronomy},
         year = 2022,
        month = mar,
       volume = {6},
        pages = {308-316},
          doi = {10.1038/s41550-022-01616-z},
archivePrefix = {arXiv},
       eprint = {2203.06199},
 primaryClass = {astro-ph.GA},
       adsurl = {https://ui.adsabs.harvard.edu/abs/2022NatAs...6..308M}
}

@ARTICLE{morishita2017,
       author = {{Morishita}, Takahiro and {Abramson}, Louis E. and {Treu}, Tommaso and {Schmidt}, Kasper B. and {Vulcani}, Benedetta and {Wang}, Xin},
        title = "{Characterizing Intracluster Light in the Hubble Frontier Fields}",
      journal = {\apj},
         year = 2017,
        month = sep,
       volume = {846},
       number = {2},
          eid = {139},
        pages = {139},
          doi = {10.3847/1538-4357/aa8403},
archivePrefix = {arXiv},
       eprint = {1610.08503},
 primaryClass = {astro-ph.GA},
       adsurl = {https://ui.adsabs.harvard.edu/abs/2017ApJ...846..139M}
}

@ARTICLE{mostoghiu2019,
       author = {{Mostoghiu}, Robert and {Knebe}, Alexander and {Cui}, Weiguang and {Pearce}, Frazer R. and {Yepes}, Gustavo and {Power}, Chris and {Dave}, Romeel and {Arth}, Alexander},
        title = "{The Three Hundred Project: The evolution of galaxy cluster density profiles}",
      journal = {\mnras},
         year = 2019,
        month = mar,
       volume = {483},
       number = {3},
        pages = {3390-3403},
          doi = {10.1093/mnras/sty3306},
archivePrefix = {arXiv},
       eprint = {1812.04009},
 primaryClass = {astro-ph.GA},
       adsurl = {https://ui.adsabs.harvard.edu/abs/2019MNRAS.483.3390M}
}

@ARTICLE{murante2007,
       author = {{Murante}, Giuseppe and {Giovalli}, Martina and {Gerhard}, Ortwin and {Arnaboldi}, Magda and {Borgani}, Stefano and {Dolag}, Klaus},
        title = "{The importance of mergers for the origin of intracluster stars in cosmological simulations of galaxy clusters}",
      journal = {\mnras},
         year = 2007,
        month = may,
       volume = {377},
       number = {1},
        pages = {2-16},
          doi = {10.1111/j.1365-2966.2007.11568.x},
archivePrefix = {arXiv},
       eprint = {astro-ph/0701925},
 primaryClass = {astro-ph},
       adsurl = {https://ui.adsabs.harvard.edu/abs/2007MNRAS.377....2M}
}

@ARTICLE{navarro1997,
       author = {{Navarro}, Julio F. and {Frenk}, Carlos S. and {White}, Simon D.~M.},
        title = "{A Universal Density Profile from Hierarchical Clustering}",
      journal = {\apj},
         year = 1997,
        month = dec,
       volume = {490},
       number = {2},
        pages = {493-508},
          doi = {10.1086/304888},
archivePrefix = {arXiv},
       eprint = {astro-ph/9611107},
 primaryClass = {astro-ph},
       adsurl = {https://ui.adsabs.harvard.edu/abs/1997ApJ...490..493N}
}

@ARTICLE{ogiya2018,
       author = {{Ogiya}, Go},
        title = "{Tidal stripping as a possible origin of the ultra diffuse galaxy lacking dark matter}",
      journal = {\mnras},
         year = 2018,
        month = oct,
       volume = {480},
       number = {1},
        pages = {L106-L110},
          doi = {10.1093/mnrasl/sly138},
archivePrefix = {arXiv},
       eprint = {1804.06421},
 primaryClass = {astro-ph.GA},
       adsurl = {https://ui.adsabs.harvard.edu/abs/2018MNRAS.480L.106O}
}


\end{document}